\documentclass{aa}

\usepackage{graphicx}
\usepackage{txfonts}
\begin{document}






   \title{Linking the evolution of terrestrial interiors and an early outgassed atmosphere to astrophysical observations}


   \author{Dan J. Bower
          \inst{1}
          \and
          Daniel Kitzmann
          \inst{1}         
          \and          
          Aaron S. Wolf
          \inst{2}
          \and
          Patrick Sanan
          \inst{3}
          \and
          Caroline Dorn
          \inst{4}
          \and
          Apurva V. Oza
          \inst{5}
          }

   \institute{Center for Space and Habitability, University of Bern,
              Gesellschaftsstrasse 6, 3012 Bern, Switzerland\\
              \email{daniel.bower@csh.unibe.ch}
         \and
             Earth and Environmental Sciences, University of Michigan,
             1100 North University Avenue, Ann Arbor, MI 48109-1005, USA\\
         \and
             Institute of Geophysics, ETH Zurich,
             Sonneggstrasse 5, 8092 Zurich, Switzerland\\
         \and
             University of Zurich, Institute of Computational Sciences,
             Winterthurerstrasse 190, 8057 Zurich, Switzerland\\
         \and
            Physics Institute, University of Bern,
            Sidlerstrasse 5, 3012 Bern, Switzerland\\
          }

   \date{Received April 16, 2019; accepted September 6, 2019}

 
  \abstract
   {A terrestrial planet is molten during formation and may remain molten due to intense insolation or tidal forces.  Observations favour the detection and characterisation of hot planets, potentially with large outgassed atmospheres.}
   {We aim to determine the radius of hot Earth-like planets with large outgassing atmospheres.  Our goal is to explore the differences between molten and solid silicate planets on the mass--radius relationship and transmission and emission spectra.}
   {An interior--atmosphere model was combined with static structure calculations to track the evolving radius of a hot rocky planet that outgasses CO$_2$ and H$_2$O.  We generated synthetic emission and transmission spectra for CO$_2$ and H$_2$O dominated atmospheres.}
   {Atmospheres dominated by CO$_2$ suppress the outgassing of H$_2$O to a greater extent than previously realised since previous studies applied an erroneous relationship between volatile mass and partial pressure.  We therefore predict that more H$_2$O can be retained by the interior during the later stages of magma ocean crystallisation.  Formation of a surface lid can tie the outgassing of H$_2$O to the efficiency of heat transport through the lid, rather than the radiative timescale of the atmosphere.  Contraction of the mantle, as it cools from molten to solid, reduces the radius by around $5\%$, which can partly be offset by the addition of a relatively light species (e.g. H$_2$O versus CO$_2$) to the atmosphere.}
   {A molten silicate mantle can increase the radius of a terrestrial planet by around 5\% compared to its solid counterpart, or equivalently account for a 13\% decrease in bulk density.  An outgassing atmosphere can perturb the total radius, according to its composition, notably the abundance of light versus heavy volatile species.  Atmospheres of terrestrial planets around M-stars that are dominated by CO$_2$ or H$_2$O can be distinguished by observing facilities with extended wavelength coverage (e.g. JWST).}


\titlerunning{Evolution of terrestrial interiors and mass--radius observations}

   \maketitle
%

\section{Introduction}
The detection of over 1000 rocky exoplanets has ushered in a new era of exoplanet characterisation, which is driven by the goal of identifying and characterising Earth-analogues or potentially habitable worlds.  During planet formation, terrestrial planets do not retain a significant primordial atmosphere from the nebula gas due to volatile loss during accretion \citep[e.g.][]{SM18}.  Rather, they outgas volatiles from the interior to form so-called secondary atmospheres once the circumstellar disk decays.  This strong coupling between the rocky interior and atmosphere presents an opportunity to probe the connection between the state of the interior and atmospheric structure and composition.  Current (TESS) and future (JWST, CHEOPS, PLATO) observatories will enhance our catalogue of rocky worlds through new discoveries and detailed characterisation, particularly of atmospheres.  These facilities demand a concurrent advancement in modelling capabilities to extract maximum insight from observational data, particularly in terms of the interplay between the interior, surface, and atmosphere.

The earliest secondary atmosphere of a rocky planet originates from extensive volatile release during one or more magma ocean epochs that occur during and after the assembly of the planet.  Magma oceans form as a result of accretion, core-formation, radioactive decay of short-lived elements, and giant impacts \citep[e.g.][]{E12}.  For example, the moon-forming impact melted at least two-thirds of Earth's mantle \citep{NS15}, producing a global magma ocean that subsequently cooled and crystallised.  Magma oceans set the stage for the long-term evolution of terrestrial planets by establishing the major chemical reservoirs of the iron core and silicate mantle, chemical stratification within the mantle, and outgassed atmosphere.  Therefore, understanding the whole life-cycle of a terrestrial planet---from its magma ocean origins to potentially a mature state with a solid mantle---has fundamental implications for the thermal and chemical evolution of terrestrial planets \citep[e.g.][]{SE18}.

Although definitions in the literature can vary, the magma ocean stage is typically defined as the period during which the planetary mantle has a melt fraction above a critical value, usually between 30\% and 50\%.  This critical value reflects the melt fraction at which a partially molten mantle abruptly switches from a dynamic regime dominated by turbulent melt transport to the comparatively slow viscous creep of solid materials.  The boundary between these two regimes is known as the `rheological transition' or the `rheological front' since it is strongly controlled by the phase and hence temperature dependency of viscosity \citep{VSS00}.  For a magma ocean that crystallises from the bottom-up, this boundary moves upwards through the mantle from the core-mantle boundary to the surface as the planet cools.

As inferred from solar system samples, the building blocks of a rocky planet can contain several hundred ppm by mass of volatiles that subsequently outgas as a magma ocean cools and crystallises.  CO$_2$ and H$_2$O are the dominant gases released from carbonaceous chondrites \citep{SF10} and potent greenhouse gases, and therefore they receive significant attention in the context of coupled interior--atmosphere evolution \citep{ET08,LMC13,NKT19}.  As a magma ocean crystallises, the solubility of volatile species in the melt phase drives outgassing as the solid fraction increases and the melt becomes over-saturated with volatiles.  The growing atmosphere thermally insulates the magma ocean and thereby dictates the early cooling rate, possibly extending the lifetime of a magma ocean from a few thousand years with no atmosphere \citep[e.g.][]{ABE93} to several tens of million years with an atmosphere \cite[e.g.][]{HAG13}.

For the terrestrial planets in the solar system, the magma ocean stage was only transient, lasting at most a few hundred million years.  However, some exoplanets may harbour permanent magma oceans, which necessitates an understanding of the spectral signature of magma oceans to assist in their characterisation.  This is because there is an observational bias to detect exoplanets with short orbital periods, such that many of the detected rocky worlds are likely to be tidally-locked and subject to intense insolation.  These factors compound to promote a permanent magma ocean on the star-facing side of the planet, which may extend globally if the atmospheric redistribution of heat is efficient \citep{kite2016atmosphere}.  Melts behave fundamentally differently than solids, which necessitates a tailored approach to determine their material properties and hence requires that their impact on interior evolution and mass--radius observations is assessed.  \cite{WB18} recently provide a thermodynamically self-consistent equation of state for MgSiO$_3$ melt---the most abundant mineral in Earth's mantle---which is calibrated by high pressure and temperature calculations and experimental data.  This raises the question of whether mass--radius data can be explained by differences in the phase of silicate matter, rather than invoking bulk compositions that depart strongly from Earth \citep[e.g.][]{DHB19}.

Magma oceans have also garnered interest from the perspective of observing planet formation in process, notably the giant impact stage, which may be amiable to direct detection using the next generation of ground-based facilities \citep{MMS09,LZM14,BLB19}.  The planet may emit strongly in the infra-red and have an outgassing atmosphere that is preferable for atmospheric characterisation.  The library of hot planets will continue to increase due to TESS preferentially detecting planets with $\sim$27 day period \cite[e.g.][]{BPQ18}, as well as the expected yield from future missions such as PLATO \citep{RCA14} and ARIEL \citep{TDE18}.  Understanding the interior evolution of a rocky planet provides a vital bridge between planet formation theory and observations.  Mass--radius data cannot uniquely constrain the structure and composition of a planet due to inherent degeneracy that is most appropriately addressed using a Bayesian framework \citep[e.g.][]{DBR18}.  Additional data or modelling constraints can be used to limit the range of acceptable interior structures, such as a combination of stellar abundances \citep{DHV16} and disk evolution models \citep{DHB19}.  Ordinarily, static structure models assume an interior thermal structure, although this is adopted a priori rather than relating directly to evolutionary considerations.  Planetary radii determined from static structure models that consider only solid phases are relatively insensitive to temperature \citep[e.g.][]{DBR18}.  However, temperature variations drive dynamic processes such as melting that facilitate outgassing, which again provides a direct connection between the interior and potentially observable atmospheric signatures.

An holistic modelling strategy is required to maximise our understanding of interior--atmosphere processes as future observational facilities push the limits of exoplanetary characterisation.  In this study, we highlight the potential to combine models of coupled interior--atmosphere evolution with static structure calculations and modelled atmospheric spectra (transmission and emission).  By combining these components in a common modelling framework, we acknowledge planets as dynamic entities and leverage their evolution to bridge planet formation, interior-atmosphere interaction, and observations.  By considering the earliest stage of terrestrial planet evolution---the magma ocean epoch---we investigate the consequences of large variations in temperature, material phase, and atmospheric properties on observations.  This has immediate application for characterising hot and close-in rocky planets.  Ultimately, we wish to close the gap between atmospheric modelling and interior dynamics, and thus pave the way for a new comprehension of the nature of rocky planets.

\section{Model}
\subsection{Dynamic interior model}
The composition space of exoplanetary interiors may be large \citep{BOL10}, which is confounded by fewer constraints on the properties of materials that depart strongly from those of terrestrial interest.  We therefore focus on a planet that has comparable size, composition, and structure to Earth, although our modelling framework itself is ambivalent and can accommodate diverse compositions and internal structures as desired.  We used the SPIDER code \citep{BSW18} to model the interior evolution of a rocky planet coupled to an atmosphere (Sect~\ref{sect:atmos}).  Our model parameters are based on Case BU presented in \cite{BSW18}, which tracks the thermal evolution of an Earth-like planet at 1 AU during its magma ocean stage.  We consider a mantle with one component (Mg-Bridgmanite) and two phases (melt and solid), which resides above an Earth-sized iron core.  The thermophysical properties of silicate melt are provided by \cite{WB18}, since they are calibrated to terrestrial mantle pressures and temperatures and agree with shock-wave data \citep[Fig.~3,][]{WB18}.  Solid properties are given by \cite{M09}, and the properties of the melt--solid mixture are derived by assuming volume additivity and computing the two-phase adiabat \citep[e.g.][]{SOLO07}.  For the melting region, we join together liquidus and solidus data appropriate for the upper and lower mantle with a smoothed transition region. The lower mantle liquidus and solidus are derived from measurements on chondritic composition \citep{ABL11}.  These melting curves have nearly linear boundaries that are reasonably described with a constant proportionality of 1.09, which is consistent with typical melting behaviour governed by the cryoscopic equation as discussed in \cite{SKS09}. For the upper mantle, we adopt the melting bounds used by \cite{HAG13} from published experimental peridotite melting data, which is likewise consistent with similar fitted boundaries \citep[e.g.][]{SKS09}.

Compared to Case BU in \cite{BSW18}, we omitted the parameterised ultra-thin thermal boundary layer at the surface; this is simply a choice to ensure that the surface temperature is resolved by both the interior and the atmosphere model.  The mesh resolution is set by 1000 nodes across the mantle.  Our model includes internal heating due to the long-lived radiogenic isotopes ($^{40}$K, $^{232}$Th, $^{235}$U, $^{238}$U), where the half-lives and present-day fractional abundances are provided by \cite{RUE17}.  We assume the mantle is undepleted (fertile) to estimate the heating rate per unit mass \citep[p. 170,][]{TS14}.  We exclude short-lived radioisotopes such as $^{26}$Al and $^{60}$Fe since our model begins as a fully assembled albeit molten planet.  Short-lived radionuclides can drive the thermal and chemical evolution of planetary building blocks \citep[e.g.][]{LGG16}, but their influence is muted during the final accretion stage.  Cases that include the gravitational separation flux driven by the density difference between the melt and solid phase have a suffix of `s' (Sect.~\ref{sect:separation}).

Our parameterised model of convection requires a prescription of the mixing length, which is conventionally taken as either a constant or the distance to the nearest boundary \citep[e.g.][]{ABE95,SC97}.  We use both of these prescriptions in this study.  A constant mixing length effectively averages the convective heat transport over the mantle and therefore replicates the results of boundary layer theory that considers an average (representative) viscosity for the calculation of the Rayleigh number.  By contrast, a mixing length that increases away from the upper and lower boundaries enforces a conductive region at the interfaces of the mantle with the atmosphere and core.  For the atmosphere interface, this can enable the near surface to form a viscous lid when it drops below the rheological transition and hence restrict subsequent cooling of the interior.  Therefore, a variable versus constant mixing length gives end-member models of slow and fast interior cooling, respectively, once the surface reaches the rheological transition.  We suffix the case number with  `v' to denote cases with a variable mixing length, that is a mixing length that increases linearly from zero at the outer boundaries to a maximum of $d/2$ in the centre of the domain, where $d$ is the thickness of the mantle.  By comparison, a suffix of `c' represents a constant mixing length of $d/4$, which is the average across the domain for the variable mixing length parameterisation.
\subsection{Coupled interior--atmosphere evolution}
\label{sect:atmos}
We modified the SPIDER code to include volatile species (H$_2$O and CO$_2$) that form an atmosphere as a consequence of outgassing as a magma ocean cools and crystallises.  The influence of radiation trapping of greenhouse species in the atmosphere can delay cooling of the interior by several orders of magnitude relative to no atmosphere being present.  For each species, mass conservation determines its abundance in the melt, solid, and gas phase, according to solubility and partitioning behaviour \cite[e.g. Eqs.~14--17,][]{LMC13}.  The partitioning of a species between the melt and solid phase is given by a constant partition coefficient, and the partitioning between the melt and gas phase is determined by a modified Henry's law that accommodates a power-law relationship between partial pressure $p_v$ and volatile abundance in the melt $X_v$:
\begin{equation}
p_v(X_{\rm v}) = \left( \frac{X_{\rm v}}{\alpha_v} \right)^{\beta_v}\,,
\label{eq:partial}
\end{equation}
where $X_v$ is the abundance of volatile $v$ in the melt, $\alpha_v$ is Henry's constant, and $\beta_v$ facilitates a non-linear relationship.  Table~\ref{table:volatile} provides the parameters for the two volatile species that we consider, H$_2$O and CO$_2$.
\begin{table}
\caption{Volatile parameters for coupled atmosphere model.}             
\label{table:volatile}      
\centering                          
\begin{tabular}{l l l}        
\hline\hline                 
Parameter & Value and units & References\\    
\hline                        
   Ref. pressure $P_0$ & 101325 Pa & \\
   CO$_2$ absorption \\~~coefficient at $P_0$ & 0.05 m$^2$/kg & 1\\      
   CO$_2$ Henry $\alpha$ & 4.4$\times10^{-6}$ ppm/Pa & 2 \& Eq.~\ref{eq:partial}\\
   CO$_2$ exponent $\beta$ & 1.0 & 2 \& Eq.~\ref{eq:partial}\\
   CO$_2$ solid--melt\\
   ~~distribution coeff. & 5$\times10^{-4}$ & 3\\
   H$_2$O absorption \\~~coefficient at $P_0$ & 0.01 m$^2$/kg & 4\\      
   H$_2$O Henry $\alpha$ & 6.8$\times10^{-2}$ (ppm/Pa)$^{1/\beta}$ & 5 \& Eq.~\ref{eq:partial}\\
   H$_2$O exponent $\beta$ & 1/0.7 & 5 \& Eq.~\ref{eq:partial}\\
   H$_2$O solid--melt\\
   ~~distribution coeff. & 1$\times10^{-4}$ & 6\\
\hline                                   
\end{tabular}
\tablefoot{References.  (1) \cite{PN03}; (2) \cite{PHH91}; (3) \cite{SWF06}; (4) \cite{YO52}; (5) \cite{CHM94}; (6) \cite{BKR03}.} 
\end{table}
The partial pressure of each volatile species relates to its atmospheric mass \citep[e.g.][]{P11}:
\begin{equation}
m_v^g = 4 \pi R_p^2 \left( \frac{\mu_v}{\bar{\mu}} \right) \frac{p_v}{g}\,,
\label{eq:pmass}
\end{equation} 
where $m_v^g$ is the mass of the volatile in the atmosphere, $R_p$ radius of the planetary surface, $\mu_v$ molar mass of the volatile, $\bar{\mu}$ mean molar mass of the atmosphere, $p_v$ the (surface) partial pressure of the volatile, and $g$ gravity.  The ratio of molar masses is omitted in some previous studies that focus on magma ocean outgassing, although the reasoning for this is unclear and difficult to justify.  The ratio of molar masses is unity for an atmosphere with a single species, but this is not the case for a multi-species atmosphere (Appendix~\ref{app:volmass}).  The mass balance equations for the volatile species form a set of coupled differential equations that we solve as part of the same system of equations that determine the evolution of the interior.  Hence our interior and atmosphere are fully coupled during the evolution.  Our atmosphere model solves for the radiative transfer of heat in a plane-parallel, grey atmosphere, with no scattering, using the two-stream approximation \citep{AM85}.  It computes an effective emissivity of the atmosphere by summing the optical depth of each volatile at the planetary surface.  The optical depth depends on an absorption coefficient (equivalent to the extinction coefficient, since scattering is not considered) that scales linearly with atmospheric pressure.  The model provides the pressure--temperature structure of the atmosphere that is fully consistent with the pressure-temperature structure of the interior.

Previous magma ocean studies often assume an initial abundance of volatiles in the (molten) mantle around 0.05 wt\% H$_2$O and 0.01 wt\% CO$_2$ \citep[e.g.][]{ET08,LMC13,NKT19}.  These values roughly correspond to the water content of Earth's ocean (300 bar) and the CO$_2$ content of Venus's atmosphere (100 bar), and are thus a lower estimate of the initial reservoir sizes that are required to explain solar system observations.  The aforementioned conversion from volatile mass to atmospheric pressure assumes that the given volatile is the only species in the atmosphere (i.e. the ratio of molar masses in Eq.~\ref{eq:pmass} is unity).  For this study, however, we are interested in the diversity of CO$_2$ and H$_2$O atmospheres that may be viable beyond the solar system.  We calibrated our initial condition to produce exactly 220 bar (surface partial pressure) of H$_2$O once all of the volatiles have outgassed, which corresponds to the pressure at the critical point of water.  For cases with a constant mixing length, we can chart the thermal evolution of the planet without H$_2$O condensing from the gas phase below the critical temperature.  For cases with a variable mixing length, the surface cools rapidly once it reaches the rheological transition and thus condensation would occur.  Since we do not model condensation we are therefore overestimating the greenhouse effect of the atmosphere for these cases.  However, the planetary cooling timescale is controlled by melt--solid separation and then viscous transport of the largely solid mantle once the rheological transition reaches the surface, rather than the atmospheric structure that determines the energy transport by radiation.  Nevertheless, the presence of water at the surface may be integral to establishing the style of convection in the viscous mantle through the initiation and sustenance of plate tectonics \citep{KJUN10}.  We consider a range of CO$_2$ volume mixing ratios at the end of outgassing:
\begin{equation}
r_{f,CO_2}= \frac{p_{f,CO_2}}{p_{f,CO_2}+p_{f,H_2O}}\,,
\end{equation}
where $r_{f,CO_2}$ is the final (i.e. fully outgassed) volume mixing ratio of CO$_2$, $p_{f,CO_2}$ the final partial pressure of CO$_2$, $p_{f,H_2O}$ the final partial pressure of H$_2$O, and the total pressure is the sum of the partial pressures according to Dalton's law.  The partial pressure of each volatile decreases with altitude above the planetary surface, but the volume mixing ratio stays the same if the atmosphere is well-mixed and condensation is neglected.  From an implementation perspective, only the partial pressure at the surface is necessary to compute for the interior coupling ($p_v$, Eq.~\ref{eq:pmass}).  We explore the influence of relatively more or less CO$_2$ compared to 220 bar of H$_2$O at the end of outgassing by varying the final volume mixing ratio of CO$_2$ (Table~\ref{table:models}).  This also results in different total surface pressures and volatile masses in the atmosphere when crystallisation is complete.  We then back compute the initial reservoir size (ppm by mass) of both H$_2$O and CO$_2$ in the fully molten mantle (Table~\ref{table:models}).  Again, even though the final partial pressure of H$_2$O is fixed in all cases to 220 bar, the initial mantle abundance of H$_2$O is not the same in all cases since CO$_2$ affects the average molar mass of the atmosphere (Eq.~\ref{eq:pmass} \& Appendix~\ref{app:volmass}).
\begin{table}
\caption{Initial volatile concentrations $X_v^0$ (ppm by mass) of CO$_2$ and H$_2$O in the mantle relative to a total mantle mass of $4.208\times10^{24}$ kg.}             
\label{table:models}      
\centering                          
\begin{tabular}{l l r r r}        
\hline\hline                 
Case & \multicolumn{2}{c}{Final outgassed} & \multicolumn{2}{l}{Initial abundance}\\
& $r_{f,CO_2}$ & $p_{f,CO_2}$ & $X_{CO_2}^0$ & $X_{H_2O}^0$ \\    
&&(bar)&(ppm)&(ppm) \\
\hline                        
1 & 0.1 & 25 & 63 & 234\\
2 & 0.13 & 32 & 79 & 227\\
3 & 0.25  & 74 & 160 & 197\\
4 & 0.33 & 110 & 221 & 181\\
5 & 0.5 & 220 & 380 & 155\\
6 & 0.66 & 441 & 666 & 136\\
7 & 0.75 & 662 & 941 & 128\\
8 & 0.83 & 1103 & 1483 & 121\\  
9 & 0.9 & 1986 & 2558 & 116\\
\hline                                   
\end{tabular}
\tablefoot{Final volume mixing ratio and partial pressure of CO$_2$ is $r_{f,CO_2}$ and $p_{f,CO_2}$, respectively.  All parameter combinations result in a 220 bar partial pressure atmosphere of H$_2$O once solidification and hence outgassing is complete.  A suffix of `v' and `c' to the case number denotes a variable and constant mixing length, respectively.  An additional suffix of `s' denotes that the case includes gravitational separation of the melt and solid phase.}
\end{table}
\subsection{Static structure calculation}
\label{sect:static}
The interior--atmosphere model determines the temperature and phase throughout the interior and the pressure--temperature structure and composition of the atmosphere.  At a given time, the mantle density is sensitive to the interior pressure and temperature, both of which also determine the phase (melt, solid, or mixed) via the melting curves.  Using this interior density distribution, we solved the static structure equations following the approach outlined in \cite{VSO07}.  The algorithm iterates to find the planetary radius of the surface that satisfies the requirement of a hydrostatic interior.  The geophysical (iron) core mass and radius are fixed to Earth values to enable us to focus on the influence of the silicate mantle and its early outgassing.  In this regard, the core is always treated as an incompressible sphere of iron at the planetary centre.  The growth of an outgassing atmosphere does not influence the density structure of the interior because the bulk modulus of silicate melt and solid is around 100 GPa.  Therefore, silicates do not experience appreciable compression due to atmospheric pressure.  We are thus justified in computing the height of the atmosphere above the planetary surface independent from the structure of the rocky interior.  The pressure--temperature structure of the atmosphere is provided by the atmosphere model \citep{AM85}.  This is converted to height--temperature using the ideal gas law (with the mean molar mass of the atmosphere) and the equation of hydrostatic equilibrium, assuming that the volatiles are well-mixed throughout the atmosphere.
\subsection{Transmission and emission spectroscopy}
We computed synthetic transmission and emission spectra at fixed times during the evolution.  The previous modelling steps provide all the information that we need for this calculation, that is the pressure--temperature structure of the atmosphere, radius of the planetary surface, and abundance of volatiles in the atmosphere.  To generate the transmission and emission spectra, we employed our observation simulator \texttt{Helios-o}. For transmission spectra we divided the atmosphere into 200 annuli $p$. For each wavenumber $\nu$ we then calculated the integrated slant optical depth along each tangent height $\tau_\nu(p)$:
\begin{equation}
 \tau_\nu(p) = \int_{-\infty}^{\infty} \chi_\nu(x(p)) \mathrm \ dx\,,
\end{equation}
where $\chi_\nu(x(p))$ is the extinction coefficient and $x$ the spatial coordinate along the tangent. The extinction coefficient contains contributions of both scattering and absorption.  The transmission $T_\nu(p)$ along a tangent $p$ is
\begin{equation}
  T_\nu(p) = e^{-\tau_\nu(p)}\,.
\end{equation}
The effective tangent height $h_\nu$ is
\begin{equation}
  h_\nu = \int_0^{\infty} \left[ 1 - T_\nu(p) \right] \mathrm \ dp
\end{equation}
and the wavelength-dependent radius $R_\nu$ is then given by
\begin{equation}
  R_\nu = R_p + h_\nu\,,
\end{equation}
where $R_p$ is the planet's surface radius, as before.  Due to the presence of a fixed surface radius and pressure, transmission spectra of terrestrial planets do not suffer from radius--pressure degeneracy \citep{Heng2017MNRAS.470.2972H}.  Synthetic emission spectra were calculated by solving the radiative transfer equation at each wavenumber $\nu$.  The spectral fluxes $F_\nu$ were obtained by using a general $n$-stream discrete ordinate method, where we employed the C version of the state-of-art discrete ordinate solver \texttt{DISORT} \citep{Hamre2013AIPC.1531..923H}.  We set the number of streams to four, which is more than sufficient to provide accurate fluxes in the absence of strongly scattering aerosols. The discrete ordinate solver takes into account absorption, thermal emission, as well as Rayleigh scattering by molecules.  The absorption coefficients were calculated using the opacity calculator \texttt{Helios-k} \citep{Grimm2015ApJ...808..182G}. For CO$_2$, HITEMP \citep{Rothman2010JQSRT.111.2139R} was employed as the source for the line list, whilst for water we used the full BT2 list \citep{Barber2006MNRAS.368.1087B}. The calculation and data sources for the molecular Rayleigh scattering coefficients are discussed in \citet{Kitzmann2017A&A...600A.111K}.
\section{Results}
   \begin{figure}
   \centering
   \includegraphics[width=\hsize]{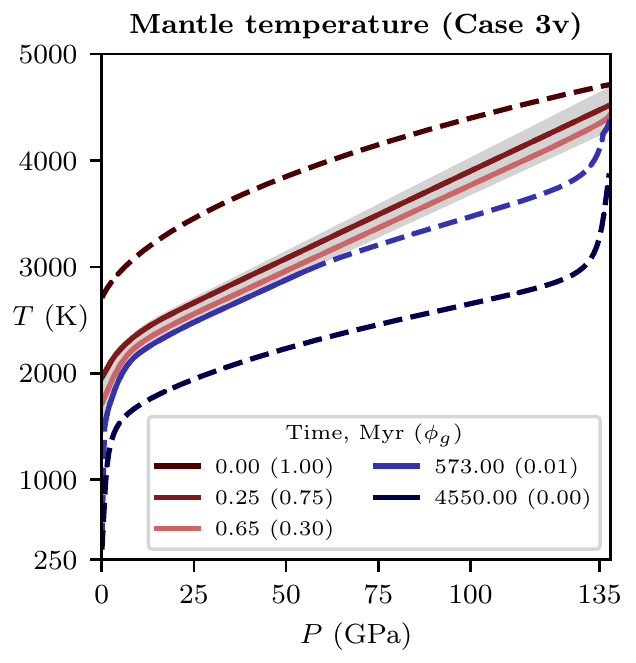}
      \caption{Evolution of interior temperature of an Earth-like planet as it cools from fully molten to fully solid (Case 3v).  Global melt fraction $\phi_g$ is given in brackets after the time in Myr.  Surface and core-mantle boundary are at 0 and 135 GPa, respectively, and the mantle is initially superliquidus (0 Myr).  Grey shaded region shows where the melt and solid phase coexist and is bounded by the liquidus above and the solidus below.  Dashed lines denote pure solid or melt, and solid lines show where solid and melt coexist.}
         \label{fig:interior}
   \end{figure}
   \begin{figure*}
      \includegraphics[width=\hsize]{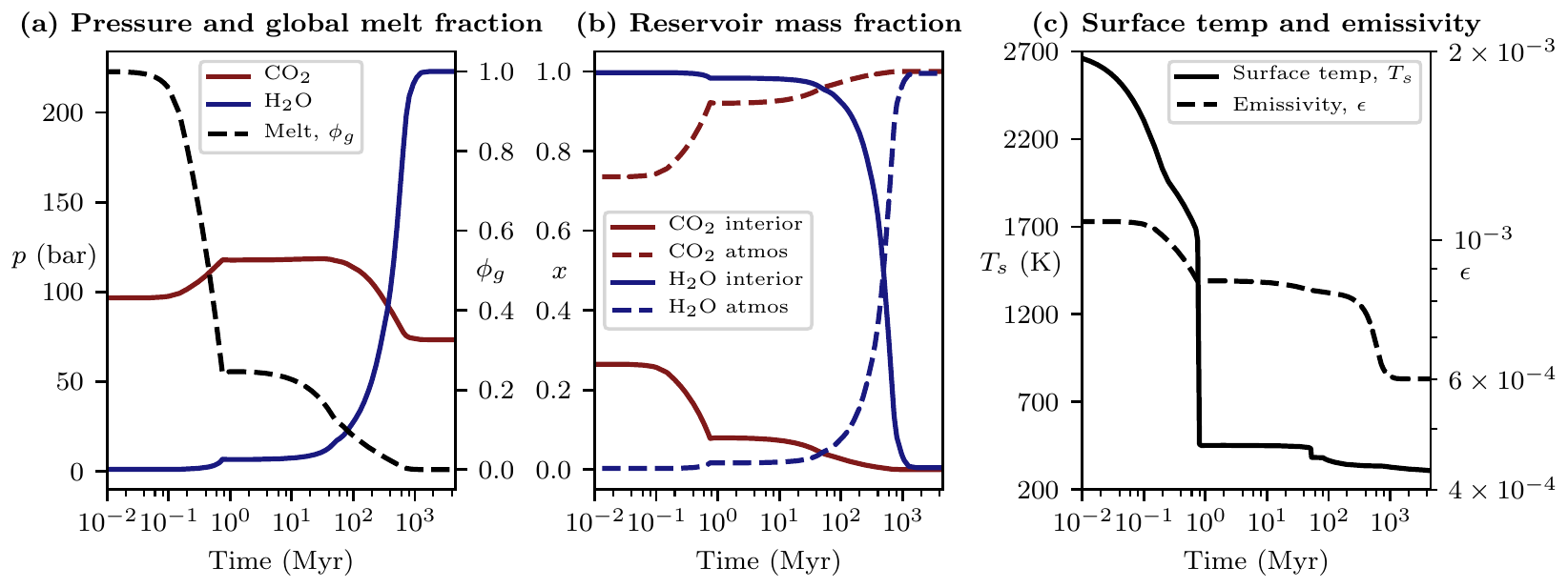}
      \caption{Evolution of atmosphere (Case 3v).  (a) Partial pressure of CO$_2$ and H$_2$O in atmosphere, and global melt fraction, (b) Volatile mass fraction in interior (melt and solid) and atmosphere, (c) Surface temperature and emissivity.}
         \label{fig:atmosphere}
   \end{figure*}
\subsection{Evolution of extended outgassing cases}
\label{sect:extended}
We first describe Case 3v (Figs.~\ref{fig:interior} and \ref{fig:atmosphere}), which encapsulates both the short- and long-term evolution of a rocky planet and demonstrates the major trends of the cases that facilitate viscous lid formation.  Case 3v uses a variable mixing length and $r_{f,CO_2}=0.25$, which produces the same final volume mixing ratio at the end of outgassing as an atmosphere with 100 bar CO$_2$ and 300 bar H$_2$O.  In previous studies, these partial pressures motivate reference cases \cite[e.g.][]{LMC13,SMD17,NKT19}, although their initial volatile abundances in the mantle are incompatible with the multi-species mass balance (Eq.~\ref{eq:pmass}).  Figures~\ref{fig:interior} and \ref{fig:atmosphere} show the coupled evolution of the interior and atmosphere as the planet cools from fully molten to fully solid.  The molten planet follows a thermal profile as dictated by the MgSiO$_3$ melt adiabat \citep{WB18}, before intercepting the liquidus at the base of the mantle.  The cooling profile then transitions through the mixed phase region, taking account of the two-phase adiabat as well as turbulent mixing of the melt and solid phase that transports energy via latent heat (Fig.~\ref{fig:interior}).  During the early cooling stage (before 1 Myr), the cooling rate is restricted by the radiative timescale of the atmosphere, which is largely determined by CO$_2$ since it is less soluble in silicate melt and hence outgasses first to the atmosphere.  The mass fraction of CO$_2$ in the interior continues to decrease during this stage and the atmosphere shows a commensurate increase (Fig.~\ref{fig:atmosphere}b).

Once the rheological transition reaches the surface, just prior to 1 Myr, there is an abrupt change in the evolutionary trajectory.  The cooling rate is now no longer restricted by the ability of the atmosphere to radiate heat, but rather the ability of the lid to transport energy to the near surface to be radiated.  At this time, the majority of H$_2$O remains in the interior reservoir (Fig.~\ref{fig:atmosphere}b), and therefore has only negligible influence on the cooling of the planet prior to this time compared to CO$_2$.  Subsequently, the surface cools rapidly to just above the equilibrium temperature (Fig.~\ref{fig:atmosphere}c), whilst the interior temperature remains below the rheological transition but above the solidus for several hundred Myr (Fig.~\ref{fig:interior}).  The volatile reservoirs do not experience significant modification during the protracted stage of cooling from around 1 Myr to 100 Myr.  After around 100 Myr, the global melt fraction $\phi_g$ drops below 10\% and H$_2$O begins to enter the atmosphere in earnest (Fig.~\ref{fig:atmosphere}a).  In our models the rheological transition is defined by a dynamic criteria in which the convective regime switches from inviscid to viscous, and not simply by a critical melt fraction or viscosity.

As H$_2$O flushes into the atmosphere, the partial pressure of CO$_2$ decreases since the volume mixing ratio of CO$_2$ decreases (Fig.~\ref{fig:atmosphere}a).  H$_2$O outgasses into the atmosphere from around 100 Myr to 1 Gyr because it is controlled by heat transport through the lid and the cooling rate of the viscous mantle.  Outgassing from the interior hence persists for around 1 Gyr and results in a 220 bar H$_2$O atmosphere and a 74 bar CO$_2$ atmosphere after 4.55 Gyr.  Of course, the initial volatile mass in the mantle was set by this criteria, but this result remains useful since it demonstrates the accuracy and stability of our fully-coupled numerical scheme to handle the interior--atmosphere evolution over short and long timescales.  For comparison, at 0 Myr the atmosphere consists of 96 bar of CO$_2$ and less than 1 bar of H$_2$O.  The thermal profile after 4.55 Gyr captures the first-order characteristics of Earth's geotherm, including average mantle temperature, core-mantle boundary temperature, and viscous boundary layers at the surface and core-mantle boundary with temperature drops around 1000 K \citep[e.g.][]{TS14}.

All extended outgassing cases are characterised by a decrease in the cooling rate once the surface reaches the rheological transition ($\phi_g$ around 30\%).  At this time, the majority of H$_2$O remains in the interior reservoir and the CO$_2$ volume mixing ratio remains higher than 85\%.  This again demonstrates that the time to cool to the rheological transition is almost exclusively determined by the abundance of CO$_2$ in the atmosphere.  For cooling beyond the rheological transition, all cases show a similar reduction in the global melt fraction with time, demonstrating that heat transport across the lid is now determining the cooling rate of the interior and not the efficiency of the atmosphere to radiate heat.
\subsection{Evolution of early outgassing cases}
\label{sect:early}
   \begin{figure*}
 \centering{
      \includegraphics[width=0.75\hsize]{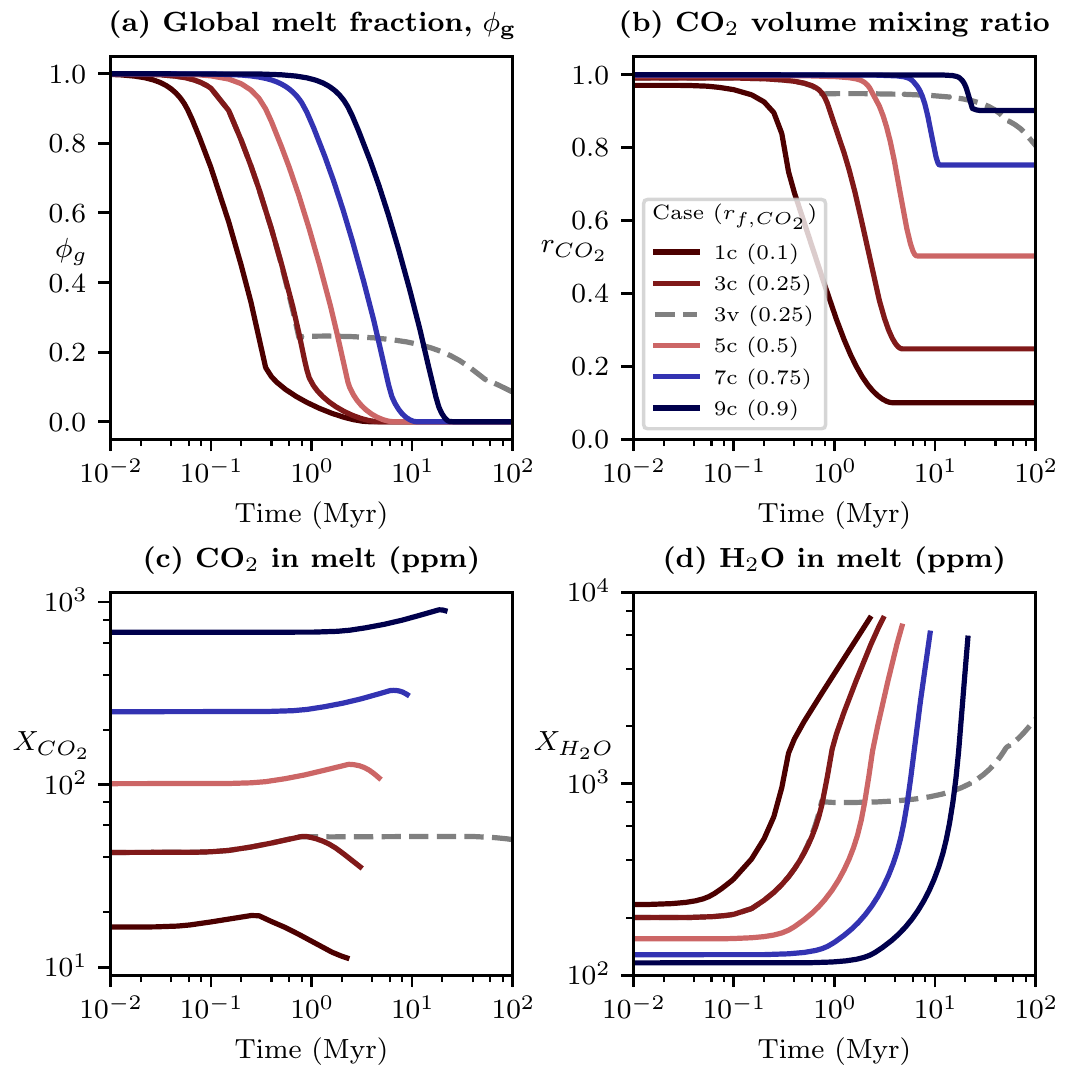}}
      \caption{Comparison of Case 1c, 3c, 3v, 5c, 7c, and 9c.  (a) Global melt fraction $\phi_g$, (b) CO$_2$ volume mixing ratio, (c) Abundance of CO$_2$ in the melt, (d) Abundance of H$_2$O in the melt.  Lines in (c) and (d) terminate at the time when the global melt fraction $\phi_g=0.01$.}
         \label{fig:all_atmosphere}
   \end{figure*}
We now present cases that demonstrate early complete outgassing of the interior.  These cases use a constant mixing length, which prevents a viscous lid from forming at the surface and hence runaway cooling of the atmosphere.  These models are therefore the most appropriate for eventual application to hot and molten exoplanets, since intense insolation maintains a molten surface and hot atmosphere and hence prevents a viscous lid from forming.  We recall that using a constant mixing length replicates the behaviour of models that use boundary layer theory.  Since our evolutionary models transition through near-equilibrium states, the mixing length parameterisation mostly determines the timescale of cooling once the rheological transition reaches the surface.  This is evident by comparing the global melt fraction $\phi_g$ for Case 3v (Figs.~\ref{fig:atmosphere}a and \ref{fig:all_atmosphere}a) with Case 3c (Fig.~\ref{fig:all_atmosphere}a).  For Case 3v, the distinctive kink in the melt fraction evolution around 1 Myr that is associated with the rheological transition reaching the surface is absent for the equivalent Case 3c with constant mixing length.

\begin{figure}
 \centering{
      \includegraphics[width=0.75\hsize]{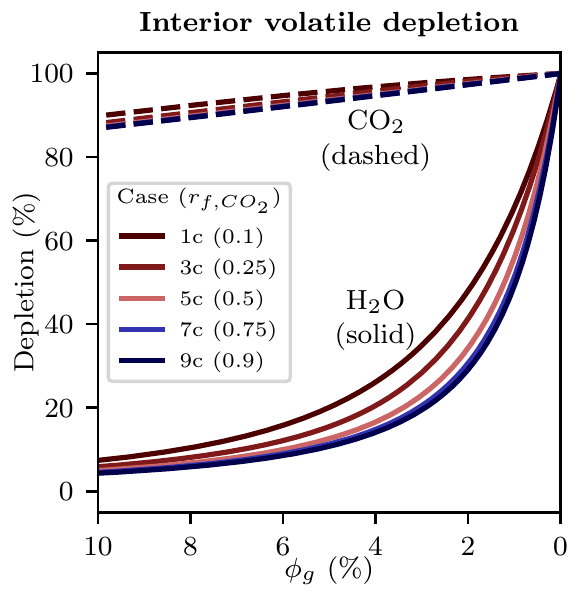}}
      \caption{Volatile depletion of interior during the later stage of crystallisation ($0\% \le \phi_g \le 10\%$) for CO$_2$ (dashed lines) and H$_2$O (solid lines).  Depletion fraction (\%) is relative to the initial amount of volatile that was partitioned into the interior according to the volatile mass balance.}
         \label{fig:all_depletion}
   \end{figure}

Figure~\ref{fig:all_atmosphere} compares the early outgassing cases.  The cooling time varies over almost two orders of magnitude, but all cases follow a similar cooling profile (Fig.~\ref{fig:all_atmosphere}a).  As the global melt fraction decreases, volatiles become over-saturated in the melt which leads to outgassing.  Outgassing of H$_2$O begins once the global melt fraction drops below about 25\% and manifests as a decrease in the CO$_2$ volume mixing ratio and an increase in the abundance of H$_2$O in the melt.  The abundance of H$_2$O in the melt always increases with time (Fig.~\ref{fig:all_atmosphere}d).  By contrast, the abundance of CO$_2$ in the melt does not always increase, and this is a consequence of the mass--partial pressure relationship (Eq.~\ref{eq:pmass}).  Following a steady increase in the melt concentration of CO$_2$, later H$_2$O outgassing causes the abundance of CO$_2$ in the melt to decrease, and this effect is most pronounced for atmospheres that are H$_2$O dominated (Fig.~\ref{fig:all_atmosphere}c).  This is because H$_2$O outgassing reduces the partial pressure of CO$_2$ in the atmosphere, such that the concentration of CO$_2$ in the melt adjusts according to Henry's law to maintain thermodynamic equilibrium (Eq.~\ref{eq:partial}).  Figure~\ref{fig:all_depletion} shows that volatile depletion of the interior occurs once the global melt fraction drops below about 15\%.  The massive CO$_2$-dominated atmospheres (e.g. Case 9c) suppress the outgassing of H$_2$O to a greater extent than atmospheres that are less massive with less CO$_2$ (e.g. Case 1c).  Whilst the total volatile mass is different for the cases, only the partial pressure of each species controls its abundance in the melt according to Henry's law (Eq.~\ref{eq:partial}).

\subsection{Influence of melt--solid separation}
\label{sect:separation}
   \begin{figure*}
      \includegraphics[width=\hsize]{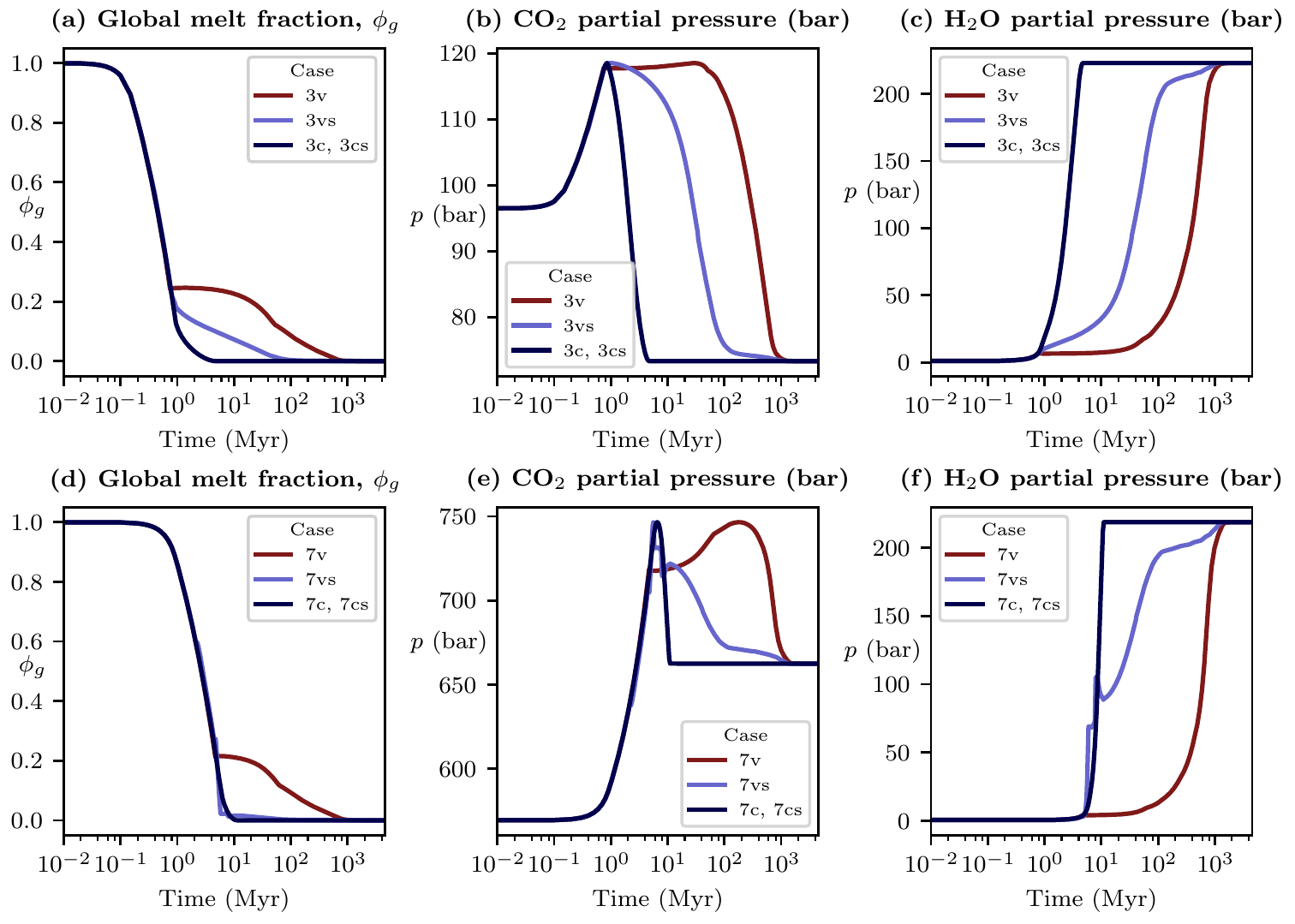}
      \caption{Evolution of atmosphere (a, b, c) Case 3, (d, e, f) Case 7.  A case suffix of `v' denotes variable mixing length, `c' constant mixing length, and `s' gravitational separation.  (a, d) Global melt fraction $\phi_g$, (b, e) Partial pressure of CO$_2$, (c, f) Partial pressure of H$_2$O.  For a constant mixing length, cases with and without gravitational separation are visually indistinguishable and therefore plotted together.}
         \label{fig:separation}
   \end{figure*}
Since we consider an Mg-Bridgmanite mantle, gravitational separation is driven by the density contrast between MgSiO$_3$ melt and solid \citep[Eq.~13,][]{BSW18}.  Therefore, the melt is on average 5\% less dense than the solid throughout the mantle according to the equations of state for MgSiO$_3$ that we adopt.  Although this fails to take account of upper mantle mineralogy and the evolving chemistry of the melt and solid phase during magma ocean cooling---both of which impact melt--solid separation---it is nevertheless useful to assess how melt--solid separation may modify the cooling timescale predicted by the extended and early outgassing cases.  Relative migration of melt and solid phases may become a dominant mechanism of energy transport once a magma ocean reaches the rheological transition \citep[e.g.][]{ABE93}.

To this end, we repeat the calculation of Case 3 and 7, with both variable and constant mixing length, to include gravitational separation (Fig.~\ref{fig:separation}).  As expected, the evolution of all four variants of each case are near identical during the early cooling stage when the global melt fraction is above the rheological transition ($\phi_g>30\%$).  Once the rheological transition is obtained, however, the trajectory of a given case variant can diverge from its companions.  Case 3vs (with separation) compared to Case 3v (without separation) shows that the cooling timescale is reduced by approximately an order of magnitude when melt--solid separation is considered.  We stress again that this is likely a maximum decrease in the cooling timescale, since melts are not expected to be buoyant at all mantle depths.  Case 3cs (with separation) and 3c (without separation) follow a near-identical evolution, such that they have been plotted as a single line in Fig.~\ref{fig:separation}.  This indicates that the gravitational separation flux is a negligible modifier to the overall heat transport for cases with a constant mixing length.  Case 7 has a larger initial CO$_2$ inventory than Case 3 which gives rise to a higher partial pressure of CO$_2$.  The general trends reported for Case 3 are also observed for Case 7; for a variable mixing length, gravitational separation decreases the cooling timescale by 1--2 orders of magnitude.  Furthermore, gravitational separation does not impact the evolution of Case 7 with a constant mixing length.
\section{Discussion}
\label{sect:discussion}
\subsection{Mass--partial pressure relationship}
\label{sect:wrong}
Several previous magma ocean studies \citep[e.g.][]{ET08,LMC13,SMD17,NKT19} do not appear to use the correct expression for the relationship between the partial pressure of a given volatile and its mass in the atmosphere (Eq.~\ref{eq:pmass}).  For example, Fig.~5 in \cite{LMC13}, Fig.~5 in \cite{SMD17}, and Fig.~4 in \cite{NKT19} show the partial pressure of both CO$_2$ and H$_2$O increasing during outgassing, whereas we find that the partial pressure of CO$_2$ decreases as the atmosphere becomes increasingly diluted by H$_2$O (Fig.~\ref{fig:atmosphere}a).  Equation~9 in \cite{ET08} also omits the ratio of the volatile molar mass to the mean atmospheric molar mass, which suggests the study is also affected by the same discrepancy.  A decrease in the CO$_2$ partial pressure is a robust feature of the parameter combinations that we investigate.  We note that studies that consider the evolution of one volatile species such as H$_2$O \citep[e.g.][]{AM85,HAG13} are unaffected.  To estimate the influence of neglecting the molar masses, we compute four extra cases: two cases using parameter combinations in this study (Sect.~\ref{sect:case37_wrong}), and two cases using an initial CO$_2$ abundance of 120 ppm and H$_2$O abundance of 410 ppm (Sect.~\ref{sect:wrong2}), which is inspired by previous reference cases \citep{LMC13,NKT19}.
\subsubsection{Influence of molar mass on Case 3c and Case 7c}
\label{sect:case37_wrong}
   \begin{figure*}
 \centering{
      \includegraphics[width=1.0\hsize]{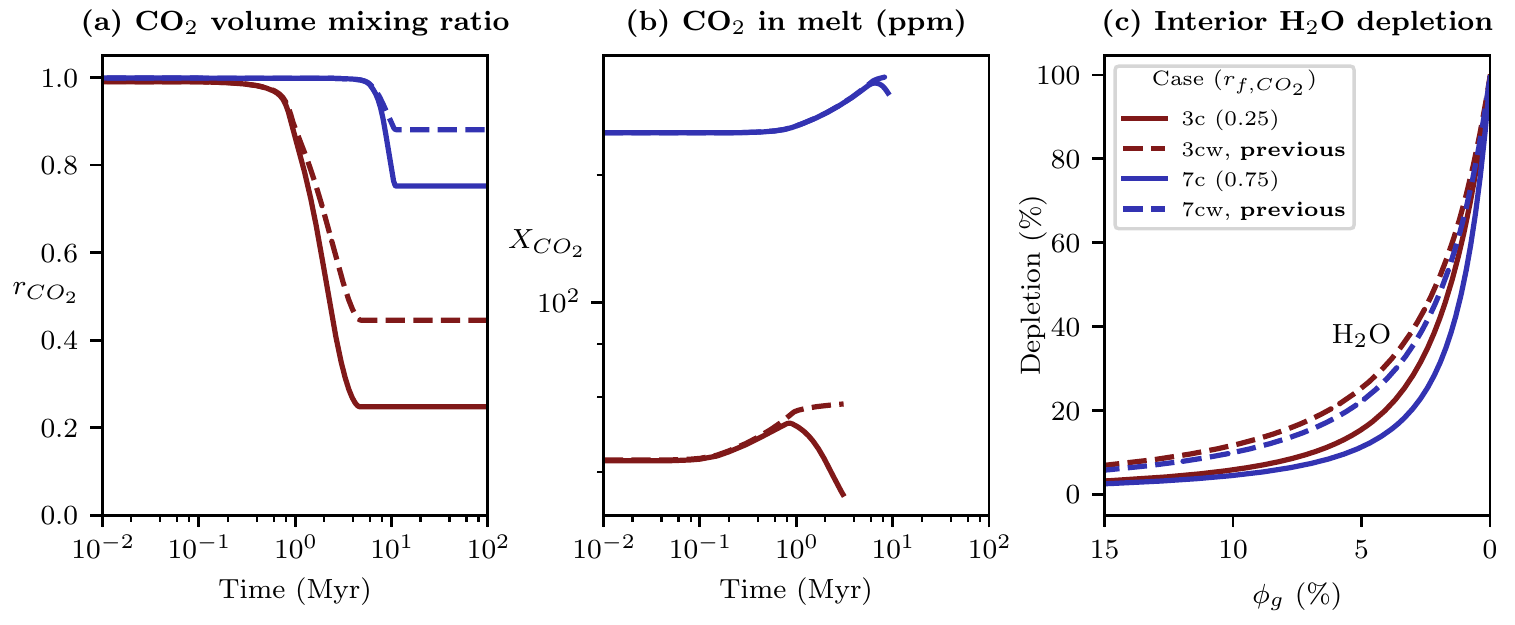}}
      \caption{Comparison of Case 3c and 7c with a previous formulation (but otherwise identical parameters) that excludes the ratio of molar masses from the volatile mass--partial pressure relationship (Eqs.~\ref{eq:pmass} and \ref{eq:volmass}).  (a) CO$_2$ volume mixing ratio, (b) Abundance of CO$_2$ in the melt, (c) Depletion fraction (\%), which is relative to the initial amount of volatile that was partitioned into the interior according to the volatile mass balance.}
         \label{fig:right_wrong}
   \end{figure*}
We compute two additional cases with exactly the same parameters as Case 3c and 7c but that omit the ratio of molar masses from the mass balance (Eqs.~\ref{eq:pmass} and \ref{eq:volmass}).  We refer to cases that include molar mass as `correct' and cases that omit molar mass as `previous formulation' or simply `previous'.  Figure~\ref{fig:right_wrong} shows the results of the correct (Case 3c and 7c) versus the previous formulation (Case 3cw and 7cw) cases.  There are three obvious differences between the correct and previous cases: (1) once H$_2$O begins outgassing, the CO$_2$ volume mixing ratio decreases at a faster rate and to a lower final value than predicted by the equivalent previous case (Fig.~\ref{fig:right_wrong}a), (2) the CO$_2$ concentration in the melt decreases at later time for the correct case, but always increases for the previous case (Fig.~\ref{fig:right_wrong}b), and (3) compared to the previous case, the correct case retains more H$_2$O in the interior until a lower global melt fraction is reached (or equivalently, until later time), but then delivery of H$_2$O to the atmosphere occurs at a faster pace.  This trade-off is because the depletion of the interior (for both correct and previous cases) begins at 0\% for a fully molten mantle and ends near 100\% for a fully solid mantle, assuming a comparatively negligible retention of volatiles in the solid mantle.  Therefore, these boundary conditions mean that the correct and previous cases have the same depletion at the start and end of evolution, but the trajectory between these two boundary values can be different.

The global melt fraction is decreasing as a consequence of continuous cooling, such that the melt volume available to accommodate the volatiles is decreasing with time.  Over-saturation of volatiles in the ever-reducing volume of melt can therefore drive outgassing.  This basic behaviour is captured by both the correct and previous cases.  The differences between the correct and previous cases can be understood by appreciating that there is feedback between the outgassing of multiple species.  For all cases, at early time the CO$_2$ volume mixing ratio of the atmosphere is near unity, since CO$_2$ is much less soluble in silicate melt than H$_2$O.  Using Eq.~\ref{eq:pmass}, we therefore find at early time that
\begin{equation}
\frac{dm_{\rm H_2O}^g}{dp_{\rm H_2O}} \approx \left( \frac{\mu_{\rm H_2O}}{\mu_{\rm CO_2}} \right) \frac{4 \pi R_p^2}{g} \approx 0.41 \times \frac{4 \pi R_p^2}{g}\,.
\end{equation}
Hence the increase of H$_2$O mass in the atmosphere for an incremental change in H$_2$O partial pressure is suppressed, compared to if only H$_2$O is in the atmosphere.  In other words, for a given addition of H$_2$O mass to the atmosphere, more solidification (cooling) is necessary to increase the abundance in the melt and hence increase the partial pressure (Eq.~\ref{eq:partial}) when H$_2$O is outgassing in the presence of CO$_2$.  Therefore CO$_2$ in the atmosphere delays the outgassing of H$_2$O until later time, and this effect is more pronounced for atmospheres that are dominated by CO$_2$ (Fig.~\ref{fig:right_wrong}c).  Relative to the previous cases, the correct cases predict a slower depletion of the H$_2$O reservoir, reaching maximum differences of 10\% at $\phi_g\approx3.5\%$ for Case 3 and 16\% at $\phi_g\approx2\%$ for Case 7.  At these global melt fractions, the surface temperature of Case 3c and 7c is 1100 K and 1400 K, respectively.  These surface temperatures are at or above the dry peridotite solidus of 1100 K \citep{H00}, and the presence of volatiles (such as CO$_2$ and H$_2$O) would further decrease the solidus temperature.  Nevertheless, the maximum differences in depletion occur when the surface is not completely molten, and in this regard the assumption of dissolution equilibrium may begin to break down.  A second mechanism explains why, also at later time, the CO$_2$ abundance in the melt decreases.  As H$_2$O flushes into the atmosphere it decreases the partial pressure of CO$_2$, and the concentration of CO$_2$ in the melt reduces to maintain thermodynamic equilibrium according to Henry's law.  CO$_2$ is always outgassing so the interior reservoir is always decreasing in total CO$_2$ mass (Fig.~\ref{fig:all_depletion}) even though its concentration in the melt can either increase or decrease.  The equations for the coupling between multiple volatile species are provided in Appendix~\ref{app:volmass}.
\subsubsection{Alternative initial volatile abundance}
\label{sect:wrong2}
   \begin{figure*}
 \centering{
      \includegraphics[width=1.0\hsize]{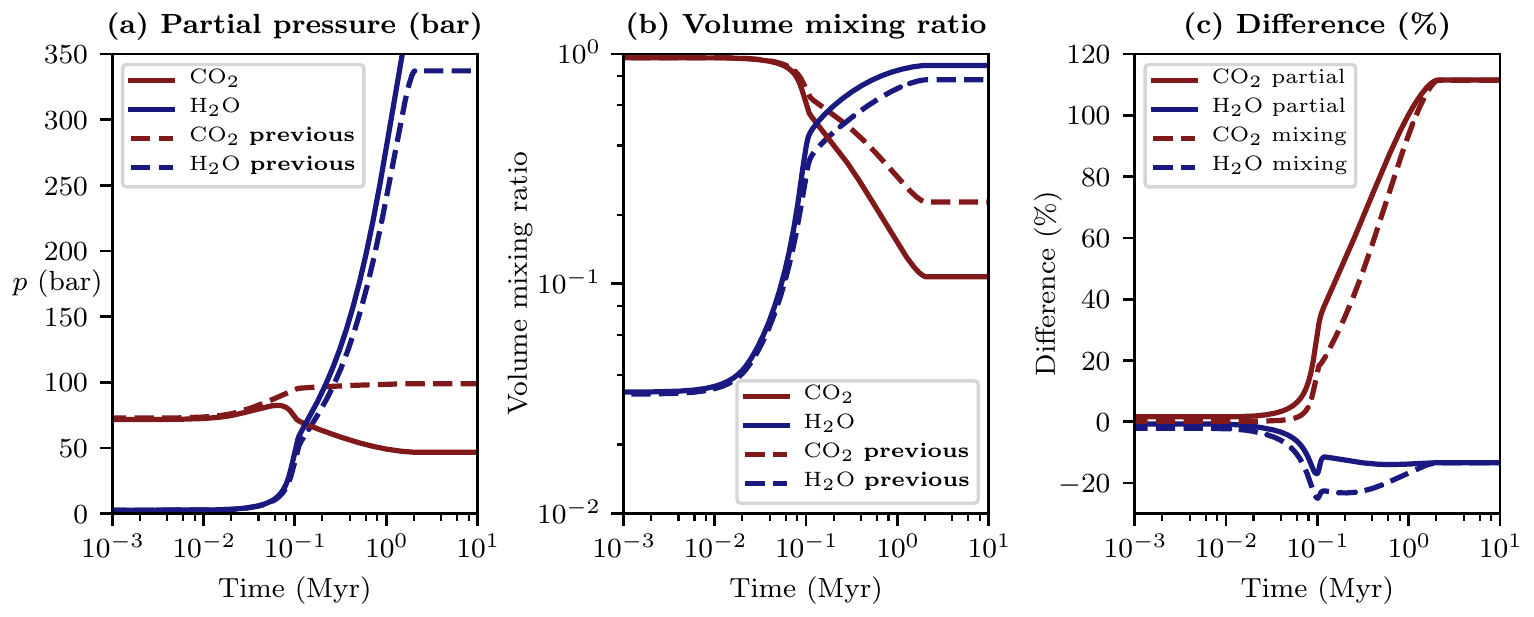}}
      \caption{Extra case with parameters similar to reference-A \citep{NKT19}.  Comparison of a correct and previous formulation case, in which the only difference is that the previous case excludes the ratio of molar masses from the volatile mass--partial pressure relationship (Eq.~\ref{eq:pmass}).  For CO$_2$ and H$_2$O: (a) Partial pressure, (b) Volume mixing ratio, (c) Relative difference of previous case to correct case (\%).}
         \label{fig:right_wrong2}
   \end{figure*}
The physical reasoning described above applies in general to two outgassing species, but to offer further calibration of existing work in the literature we compute another two cases (one correct, one previous formulation).  We configure a model based on the reference-A model in \cite{NKT19}, which itself is similar (by design) to the models in \cite{LMC13}.  This model setup is typical of magma ocean models in the literature that investigate early outgassing of volatile species.  Compared to the standard model setup for the constant mixing length cases in this study, we make the following amendments.  We exclude core cooling, internal heat generation, and the turbulent mixing of the melt and solid phase.  The initial CO$_2$ and H$_2$O are 120 and 410 ppm, respectively, and we decrease the CO$_2$ absorption coefficient to 10$^{-4}$ m$^2$/kg.  We deem our model similar enough to reference-A \citep{NKT19} to provide a meaningful assessment of how neglecting the molar mass ratio term in the mass--partial pressure relationship affects previous results.

Figure~\ref{fig:right_wrong2} shows a comparison of correct and previous cases and can be compared with Fig. 4 in \cite{NKT19}.  Both cases behave similarly before the outgassing of H$_2$O.  Outgassing of H$_2$O begins around the same time (soon after 10$^{-2}$ Myr) and the partial pressure of H$_2$O increases at a similar rate (Fig.~\ref{fig:right_wrong2}a).  Here we have kept the initial volatile abundances the same rather than the final mixing ratios at the end of outgassing, so the H$_2$O partial pressure in the correct case continues to rise even though the previous case reaches a maximum around 340 bar.  At 10 Myr, the mixing ratio for H$_2$O is less for the previous case (Fig.~\ref{fig:right_wrong2}b), resulting in a maximum error around -25\% during the evolution (Fig.~\ref{fig:right_wrong2}c).  The most pronounced difference by far is the error that accumulates for the partial pressure and volume mixing ratio of CO$_2$.  As we previously note, the partial pressure of CO$_2$ decreases as a consequence of H$_2$O outgassing, and this behaviour is again evident for these cases (Fig.~\ref{fig:right_wrong2}a).  For the previous formulation, this results in more than 100\% error for the CO$_2$ partial pressure and mixing ratio at 10 Myr.  We further confirm the trend shown in Fig.~\ref{fig:right_wrong}c that the correct case predicts less depletion of the H$_2$O interior reservoir (figure not shown).  Therefore, this preliminary analysis suggests that results of previous studies are impacted by the difference of the molar masses of CO$_2$ and H$_2$O.

\subsection{Magma ocean solidification and outgassing of H$_2$O}
\subsubsection{Magma ocean lifetime}
There are broadly two ways to define the lifetime of a magma ocean.  Firstly, the lifetime can be the time taken for either the surface or the bulk of the mantle to reach the rheological transition (`RT definition'); below this transition the cooling timescale is dictated by melt--solid separation and ultimately the viscous creep of the solid matrix.  Secondly, the lifetime may instead be the time for a magma ocean to freeze to a near or fully solid state (`solid definition').  These definitions are sensitive to different parameters and regimes of a cooling magma ocean.  The lifetime according to the RT definition is dictated by the efficiency of the atmosphere to radiate heat as well as the melting curves near the surface which determine when the RT is reached.  In this case, the cooling timescale is controlled by radiative energy emission from the hot surface and is not limited by heat transfer to the surface, which is rapid for a turbulent magma ocean.  This is why cooling to the RT is not sensitive to the choice of mixing length (variable or constant) or the inclusion of gravitational separation (Fig.~\ref{fig:separation}).  Cooling from a molten state to the RT is rapid without an insulating atmosphere, occurring from a few hundred years \citep{VSS00} to several hundred thousand years \citep{MAS16}.  This lifetime is increased by orders of magnitude to 1 Myr to 100 Myr for models that account for a co-evolving atmosphere \citep{ET08,LMC13,HAG13,NKT19}.

For the solid definition of lifetime, the atmospheric cooling rate and interior melting curves can conspire to give similar cooling times, albeit for completely different reasons.  For example, models that assume cooling is limited by radiative heat loss can crystallise 98\% of the Earth's mantle in less than 5 Myr \citep{ET08}.  We expect this time to increase if we take account of the transition to a cooling timescale dictated by a viscous lid or interior dynamics once the surface reaches the RT.  However, the formation of a viscous lid can be mitigated by upwards draining of buoyant melt through a solid matrix, which keeps the near surface hot at the expense of cooling the deep interior (Fig.~\ref{fig:separation}).  Although a dynamic model of the interior reports a similar lifetime to \cite{ET08} of a few Myr \citep{LMC13}, this is a consequence of very steep melting curves such that most of the mantle is crystallised by the time the surface reaches the RT.  \cite{LMC13} and \cite{ABE97} use melting curves based on an ideal mixture of MgSiO$_3$ and MgO extrapolated to high pressure, which results in a liquidus temperature at the base of the mantle of over 6000 K.  This is around 800 K higher than the liquidus of peridotite \citep{FAS10} and 1250 K higher than the liquidus of chondritic mantle \citep{ABL11}.  \cite{SMD17} update the model of \cite{LMC13} to use chondritic melting curves \citep{ABL11} and find that the surface temperature reaches the RT between around 500 kyr and 1 Myr.  These time estimates are impacted by a miscalculation of the outgoing long-wavelength radiation in their atmosphere model \citep[see][for discussion]{MSM17}.

The majority of previous magma ocean models use boundary layer theory to model interior dynamics, which facilitates continuous outgassing since they do not consider lid formation at the surface which can restrict cooling.  Rather, these models compute an estimate of the bulk viscosity of the mantle that is then used for calculating an effective Rayleigh number and hence the magma ocean heat flux.  We reproduce these types of models using a constant mixing length, which also has the effect of disregarding a near-surface lid that can influence the cooling rate.  We refer to these cases as early outgassing cases (Sect.~\ref{sect:early}).  However, due to the true 1-D nature of our model, we also investigate cases that form a viscous lid at the surface, resulting in extended outgassing cases (Sect.~\ref{sect:extended}).  These cases represent an end-member scenario in which melt--solid separation is insufficient to maintain a hot surface and prevent lid formation.  For these cases, there is a dramatic change in cooling style once the surface reaches the RT, since the growth of a lid greatly hinders heat exchange from the interior to the atmosphere.  The surface cools rapidly due to efficient radiation of heat and the comparatively sluggish internal (viscous) dynamics that cannot replenish the near surface with heat at the same rate due to presence of the viscous lid.  Hence the planet transitions from a cooling timescale dictated by the atmosphere to a cooling timescale dictated by the lid and the interior dynamics.  The liquidus and solidus of the mantle determine both when the surface reaches the RT and the global melt fraction at this time, which depends on the gradient of the melting curves with pressure and their relationship to the thermal profile.

Gravitational separation of melt and solid phases can reduce the cooling timescale of the extended outgassing cases by around an order of magnitude, and does not appreciably alter the cooling timescale of the early outgassing cases.  At the RT, upwards draining of buoyant melt delivers heat from deep within the mantle to the surface, which maintains a hot surface to enable both efficient radiative cooling as well as preventing a viscous lid from forming.  Therefore, the timescale for magma ocean cooling with gravitational separation is bracketed by the extended and early outgassing cases (Fig.~\ref{fig:separation}).  We can furthermore consider a melt--solid density crossover that exists at high pressure, such that melt drains downwards from a (buoyant) solid matrix to form a basal magma ocean \citep[e.g.][]{LHC07}.  For this (unmodelled) crossover scenario, we surmise that the global melt fraction would remain higher for extended time, compared to a scenario where melts are buoyant throughout the mantle.  In addition to the transport of heat downwards due to melt migration, the growth of a mid-mantle viscous layer can restrict the cooling of the basal magma ocean beneath.  This is independent of whether or not a viscous lid at the surface is also present or forming.  The mid-mantle layer may also render the assumption of dissolution equilibrium invalid with respect to the molten reservoir beneath, such that a proportion of the initial volatile budget is trapped in the deep mantle and unable to readily outgas.
\subsubsection{Outgassing volatiles}
Despite differences in methodology and the mass--partial pressure relationship (Sect.~\ref{sect:wrong}), for all cases we also find that CO$_2$ controls the early cooling rate of the interior and H$_2$O outgasses later \citep{LMC13,NKT19}.  However, this result is strongly controlled by the solubility curves, which are often the same or similar between studies, so agreement in this regard is not surprising.  A more interesting result, is that the outgassing of H$_2$O is strongly impacted by the delayed cooling and solidification of the mantle caused by a surface lid (extended outgassing cases).  For the initial H$_2$O and CO$_2$ combinations that we consider in this study, we find that the majority of H$_2$O does not outgas during the magma ocean stage prior to the surface reaching the RT.  Instead, 90\% of the initial H$_2$O reservoir (by mass) remains in the interior and its outgassing rate is determined by heat transport in the mantle (gravitational separation and viscous flow) and the surface lid.  Late or extended outgassing of H$_2$O, or any volatile for that matter, is a challenge to model since it falls outside the melt fraction regime that is typically considered by either short- or long-term modelling approaches.

When the surface reaches the RT, the interior dynamics are switching to a viscous flow regime according to the local Reynolds number and melt--solid separation can become an important energy transfer mechanism.  The transition from inviscid to viscous convection is validated by recent studies that demonstrate that solid-state convection begins during magma ocean crystallisation \citep{BLH17,MTS17}.  We thus capture the first-order nature of the transition that is important for bulk heat transport, but the details of how a lid forms and evolves at the surface is complex, notably depending on melt migration in the near-surface environment.  Nevertheless, if significant volatile outgassing from terrestrial planets occurs once the surface has partially solidified to form a lid, then there could be profound feedback between processes operating in the interior, atmosphere, and surface.  The lid-forming process is particularly relevant for H$_2$O outgassing, since H$_2$O is relatively soluble in silicate melt and does not readily outgas until the global melt fraction drops to a few tens of percent.

For a fixed partial pressure of H$_2$O at the end of outgassing, Fig.~\ref{fig:all_atmosphere}a shows that the time taken to reach the RT is approximately monotonically increasing with increasing initial CO$_2$ abundance.  This appears to contradict Fig.~6 in \cite{SMD17}, which shows the RT is first obtained for a case with an atmosphere that has a reduced initial CO$_2$ content.  \cite{SMD17} cases show a similar early cooling trajectory to our cases, with a high initial CO$_2$ abundance resulting in slower cooling and higher surface temperature relative to low initial CO$_2$ abundance.  However, the crucial difference is that whilst both studies use the same H$_2$O absorption coefficient of $10^{-2}$ m$^2$/kg, we use a CO$_2$ absorption coefficient of $5\times10^{-2}$ m$^2$/kg (i.e. larger than the H$_2$O value) whereas \cite{SMD17} use $10^{-4}$ m$^2$/kg (i.e. smaller than the H$_2$O value).  Once H$_2$O begins to outgas it dilutes the CO$_2$ in the atmosphere and drives the effective absorption coefficient of the multi-species atmosphere towards the value of the H$_2$O absorption coefficient.  For \cite{SMD17}, this means that the case with low initial CO$_2$ abundance is strongly diluted by H$_2$O and hence cooling then proceeds slower due to the influence of the relatively large H$_2$O absorption coefficient.  This is compared to the case with high initial CO$_2$ abundance that is diluted less by H$_2$O and hence retains an effective absorption coefficient that is closer to the value for pure CO$_2$.  For our cases, we do not observe such an appreciable change in cooling rate once H$_2$O outgasses because the absorption coefficients of CO$_2$ and H$_2$O are comparable---only differing by a factor of five versus a factor of 100 in \cite{SMD17}.

We apply the volatile mass balance (Eq.~\ref{eq:volmassdt}) throughout the evolution of the extended outgassing cases, which implies that the interior and atmosphere are always in thermodynamic equilibrium even once the surface cools and forms a lid.  Hence, outgassing in our models is restricted by the planetary cooling rate and not the ability of the interior to communicate with the atmosphere.  Although we compute the evolution of Case 3v for several Gyr (Fig.~\ref{fig:interior}), our model does not consider any volatile sinks, such as atmospheric escape or ingassing of volatile species due to geochemical cycling driven by surface processes such as plate tectonics.  Rather, our objective is to demonstrate the switching of timescales that control planetary cooling, and to note the potential importance of the surface lid and near-surface dynamics that has largely been ignored by previous modelling studies.

\subsection{Implications for observations}
   \begin{figure*}
 \centering{
      \includegraphics[width=0.75\hsize]{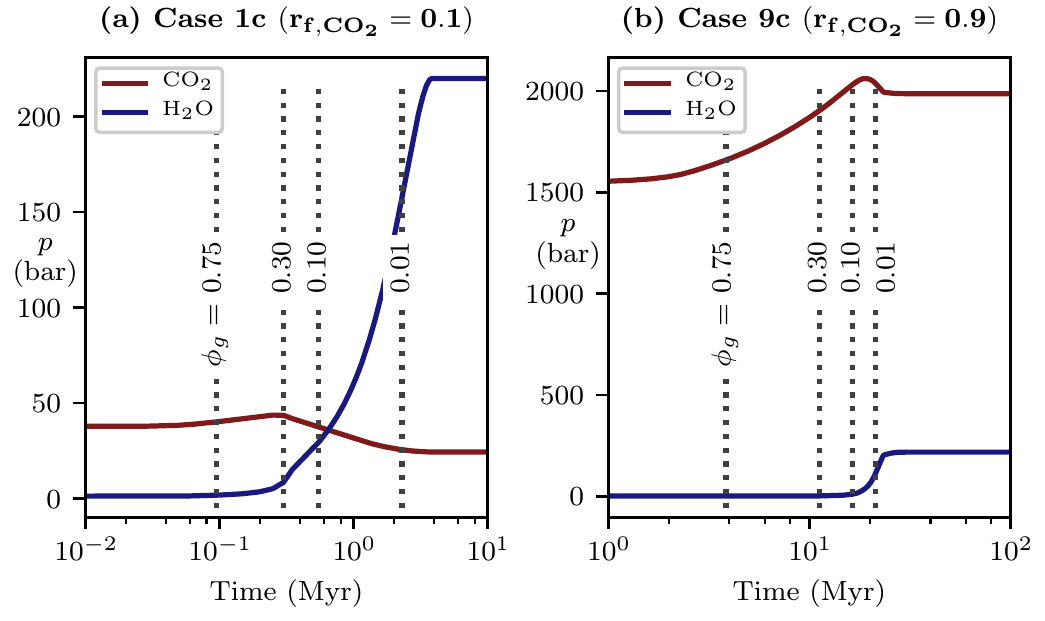}}
      \caption{Evolution of partial pressure of CO$_2$ and H$_2$O of (a) Case 1c, (b) Case 9c.  Vertical dotted lines show the times at which the interior reaches a global melt fraction $\phi_g=$ 0.75, 0.3, 0.1, and 0.01.}
         \label{fig:case1_case9_atmosphere}
   \end{figure*}
We apply the results of our evolutionary models to the detection and characterisation of rocky exoplanets by considering the global melt fraction at various stages of the evolution.  This is facilitated since our models evolve through a series of quasi-equilibrium states, which enables us to inform both observations of transient magma oceans \citep[e.g.][]{MMS09} as well as permanent magma oceans on `lava planets' that are sustained by intense insolation.  Although the mechanism by which a melt reservoir is formed and sustained is different for these two scenarios, both result in a molten or partially molten planet, and we now focus on the observable nature of such a planet.  Numerous hot terrestrial planets are expected to be discovered by current (e.g. TESS) and future (e.g. PLATO, ARIEL) observatories, and even though they exhibit extreme environments from the perspective of habitability, they remain at the cutting edge of observations.  Figure~\ref{fig:case1_case9_atmosphere} shows the evolution of the partial pressure in the atmosphere for Case 1c and Case 9c, with a relatively low ($r_{f,CO_2}=0.1$) and high ($r_{f,CO_2}=0.9$) final volume mixing ratio of CO$_2$, respectively.  We determine the global melt fraction $\phi_g$ and surface temperature ($T_s$) at several times which are denoted by vertical dotted lines (Fig.~\ref{fig:case1_case9_atmosphere}).  For these global melt fractions, we combine the integrated interior--atmosphere model with structure calculations and spectra calculations to determine the observational nature of an Earth-like planet with an early outgassed atmosphere (Fig.~\ref{fig:static}).
\subsubsection{Mass--radius}
   \begin{figure*}
 \centering{
      \includegraphics[width=0.75\hsize]{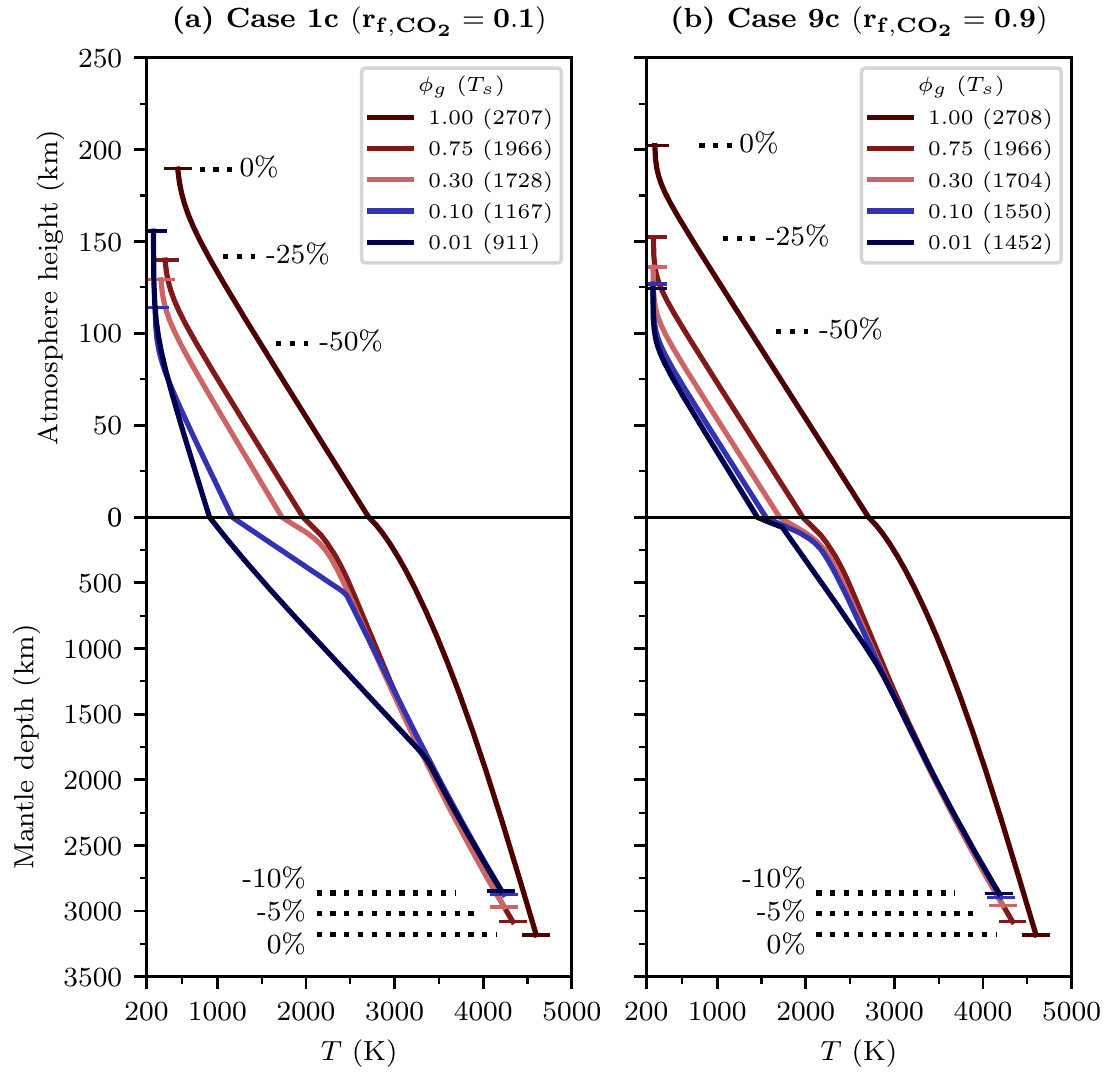}}
      \caption{Combining the integrated interior--atmosphere evolutionary model with internal structure calculations.  Global melt fraction is $\phi_g$ and surface temperature is $T_s$.  Planetary surface is at 0 km and the atmospheric height and mantle depth are plotted relative to this coordinate.  Horizontal bars at the end of each profile show the 10 mbar pressure contour (uppermost bar) and the core-mantle boundary depth (lowermost bar).  Horizontal dotted lines indicate the percentage change in height (atmosphere) and depth (mantle) relative to the initial height and depth of the respective reservoir at 0 Myr.  (a) Case 1c, (b) Case 9c.}
         \label{fig:static}
   \end{figure*}
\begin{figure*}
 \centering{
      \includegraphics[width=1.0\hsize]{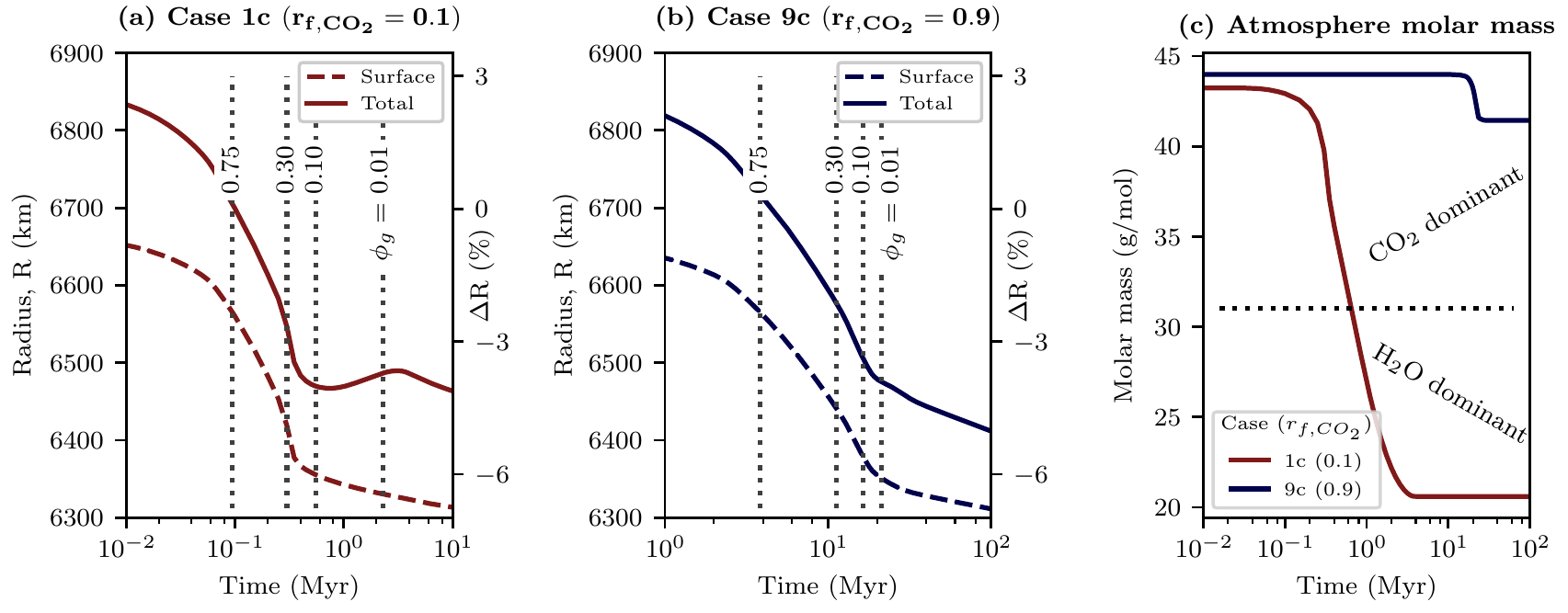}}
      \caption{Evolution of planetary surface radius and total radius (including the atmosphere, i.e. the 10 mbar contour) of (a) Case 1c, (b) Case 9c.  Left y-axis shows the radius $R$ (km) and right y-axis shows the difference relative to the planetary radius with a fully molten mantle $\Delta R$ (\%) (i.e. relative to the initial condition).  Vertical dotted lines show the times at which the interior reaches a global melt fraction of $\phi_g=$ 0.75, 0.3, 0.1, and 0.01.  (c) Mean molar mass of the atmosphere.}
         \label{fig:radius}
   \end{figure*}

Figure~\ref{fig:static} reveals that the thickness of the silicate mantle is around 10\% less for a fully solid mantle than for a fully molten mantle.  A mantle that is locked at the rheological transition ($\phi_g$ about 30\%) has a thickness around 5\% less.  The contraction of the mantle follows a similar trend for both Case 1c and 9c, which is mostly a consequence of plotting the same global melt fraction $\phi_g$.  By contrast, the atmospheric structure of the two cases exhibits crucial differences.  The massive CO$_2$ atmosphere of Case 9c undergoes continual contraction due to cooling, such that it reduces in height by around 40\% as the mantle transitions from dominantly molten ($\phi_g=1$) to solid ($\phi_g=0.01$).  Case 1c also contracts as $\phi_g$ decreases, but we notice that the atmospheric height is predicted to be higher for $\phi_g=0.01$ than for $\phi_g=0.10$.  This is because H$_2$O flushing into the atmosphere during the later stage of the evolution reduces the mean molar mass of the atmosphere such that the atmosphere expands.  This dilution of the atmosphere also occurs for Case 9c, but not to the same extent since there is insufficient H$_2$O available to appreciably decrease the mean molar mass of the atmosphere (Fig.~\ref{fig:radius}c).
   \begin{figure}
 \centering{
      \includegraphics[width=1.0\hsize]{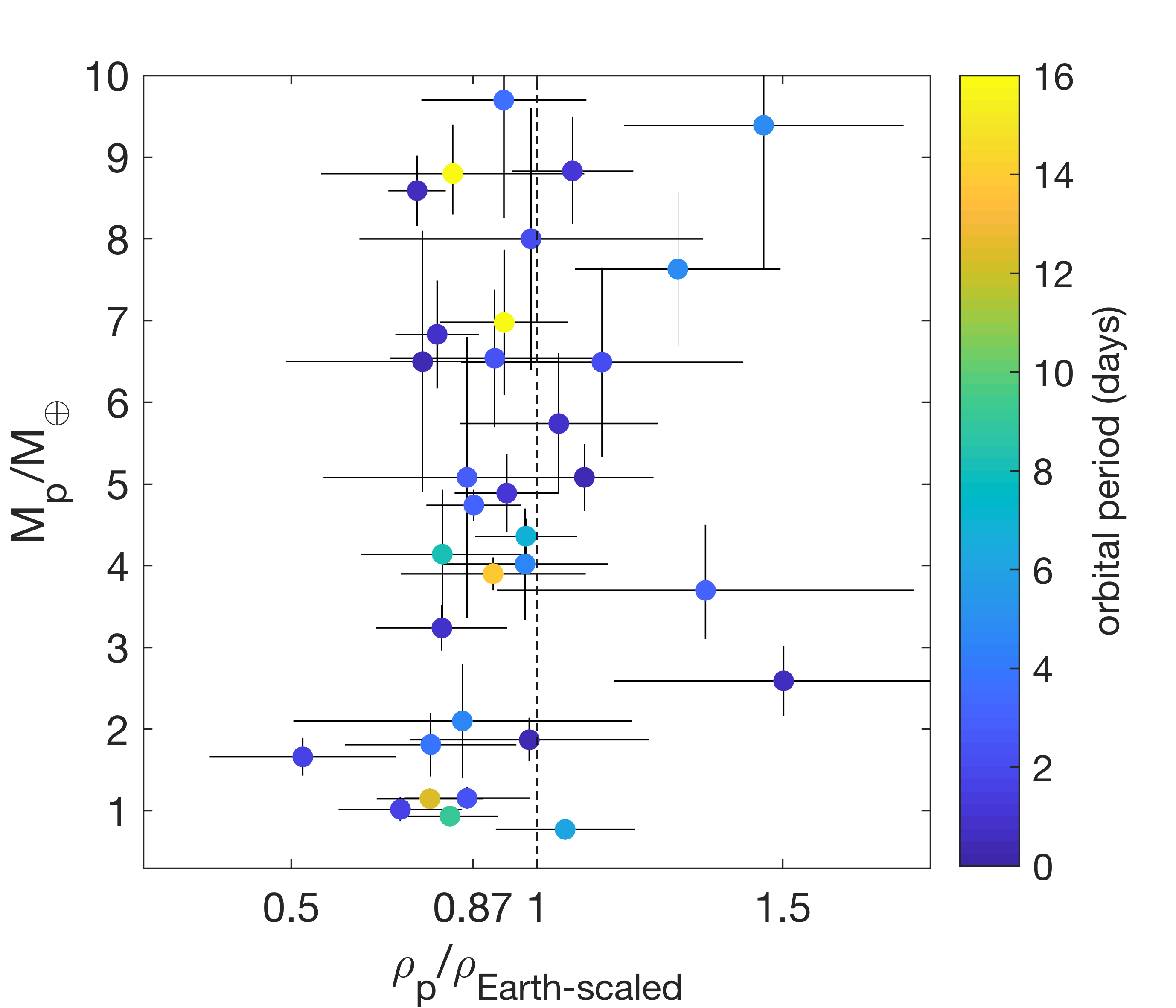}}
      \caption{Mass versus density diagram of confirmed exoplanets as a function of orbital period.  Density is relative to an Earth-like interior $\rho_{\rm Earth-scaled}$ that corresponds to a planetary mass $M_{\rm p}$.  Molten mantles alone can lower the bulk density to $\rho_{\rm p}$/$\rho_{\rm Earth-scaled}$ = 0.87.  All planets are close-in such that irradiation and tidal heating may prohibit complete solidification.  Planet sample is taken from the TEPCat catalogue (http://www.astro.keele.ac.uk/jkt/tepcat/) and shows planets with radii less than twice Earth ($R_{\rm p} < 2 R_{\oplus}$), and mass and radius uncertainties below 35\% and 15\%, respectively.}
         \label{fig:planet_mass}
   \end{figure}

We explore the connection between outgassing and planetary radius by determining the evolution of the radius for Case 1c and 9c (Fig.~\ref{fig:radius}a, b), for both the planetary surface and the total radius which includes the atmospheric height (defined at 10 mbar).  Initially the hot atmosphere increases the planetary radius relative to a molten mantle by 2\% to 3\%.  The early evolution of the radius before the rheological transition is reached is driven by contraction of the mantle with a smaller contribution from atmospheric cooling.  Beyond the rheological transition, the mantle continues to contract to reduce the total radius by a further 1\% at the end of solidification.  For the CO$_2$-dominated atmosphere (Case 9c), the total radius tracks the planetary surface radius, suggesting that contraction of the atmosphere is near complete once the rheological transition is reached.  For Case 1c, however, outgassing of H$_2$O once the planet reaches around 10\% melt fraction is able to compensate for the reduction in radius due to mantle contraction, and increases the total planetary radius before it then decreases again (Fig.~\ref{fig:radius}a).

Our results show that variations in mantle melt fraction and atmospheric thickness might be the cause of observed variability in the bulk densities of super-Earths below two Earth radii (Fig.~\ref{fig:planet_mass}).  Super-Earths do not follow a simple mass--radius trend, but rather reveal a diversity of mass--radius relationships that are usually associated with compositional and structural differences.  Close-in planets that are highly irradiated and experience strong tidal heating \citep[e.g.][]{BSR13} or induction heating due to the stellar magnetic field \citep{KNJ17}, might have largely molten interiors which reduces the bulk density similar to our calculated planetary interiors (Fig.~ \ref{fig:static}).  If the density decrease observed here for molten Earth-sized bodies is of similar scale for more massive super-Earths, a completely molten mantle could explain the radius of close-in rocky exoplanets such as HD219134 b and 55 Cnc e.  This offers an alternative explanation to these planets having exotic solid compositions with light element enrichment and possibly a mineral-rich atmosphere \citep[e.g.][]{DHB19}.

It remains at the cutting edge of material science to experimentally verify the equation of state of planetary materials at the extreme interior conditions of super-Earths \citep[e.g.][]{WSF18}.  We have intentionally limited our exposure to large uncertainty in the equation of state by focussing on a planet that is Earth-size and with Earth-like composition.  Nevertheless, we estimate that the mantle thickness can change by up to 10\% during cooling from a molten to solid state.  Assuming an Earth-like core size, this mantle decrease translates to a 13\% decrease in bulk density or a change in bulk radius of 5\% for the rocky interior.  The evolution of the coupled interior--atmosphere models indicate a change in planetary radius of up to 7\%.  Such differences in density or radii can be measured with current and future observational facilities (e.g. HARPS, TESS, CHEOPS, PLATO).  Especially with CHEOPS, well-characterised radii for a distribution of super-Earths might confirm an overall decrease in the bulk density of planets at small orbital distances and relatively low (Earth-like) planet mass.


\subsubsection{Planetary spectra}
   \begin{figure*}
 \centering{
      \includegraphics[width=0.75\hsize]{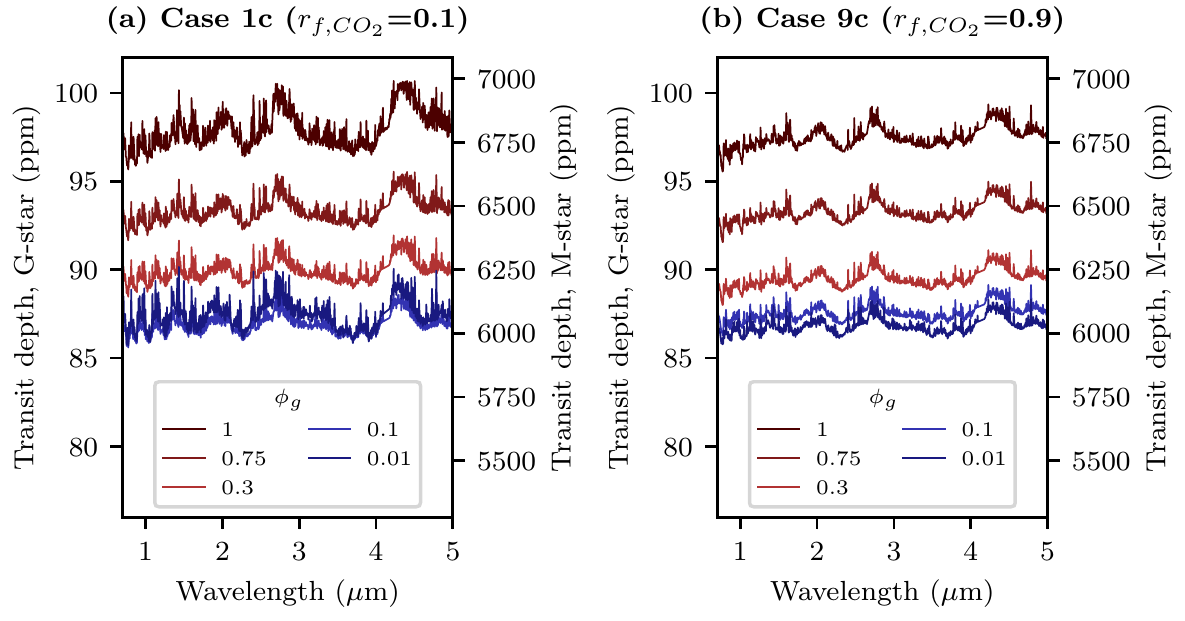}}
      \caption{Prediction of transit depths for mantle melt fraction $\phi_g=$ 1, 0.75, 0.3, 0.1, 0.01.  (a) Case 1c, (b) Case 9c.  See Fig.~\ref{fig:static} or \ref{fig:emission} to relate the global melt fraction $\phi_g$ to surface temperature $T_s$.  The left and right y-axes show the transit depth for a G-star and M-star, respectively.}
         \label{fig:transmission}
   \end{figure*}
We compute transmission and emission spectra for Case 1c and Case 9c at global melt fraction $\phi_g$=1, 0.75, 0.3, 0.1, 0.01.  These spectra correspond to the static structure calculations in Fig.~\ref{fig:static}.  For each case, the transmission spectra are calculated for two central stars: a G2V star (1 solar radius, left y-axis) and a cool M-dwarf (0.117 solar radii, right y-axis) (Fig.~\ref{fig:transmission}).  For a G-type main sequence star, the total variations in transit depth between the different melt fractions are of the order of 15 ppm for each of the two cases. Whilst the primary transits of these planets could be detected by the PLATO satellite mission \citep{RCA14}, the variations due to the different melt fractions are undetectable in the near future. It is also virtually impossible to distinguish between Case 1c and 9c with the currently planned, future planet detection and characterisation missions if the planet is orbiting a solar-like star.

By contrast, if the planet orbits a cool M-dwarf, the transit depths increase considerably (right y-axis, Fig.~\ref{fig:transmission}) due to the much smaller stellar radius. The overall variations of the transit depths are now in the range of 750 ppm, which are even detectable with current ground-based telescopes.  For the same melt fraction, Case 1c and 9c differ of the order of 100 ppm. These variations are caused by the different atmospheric compositions: Case 1c is a water-dominated atmosphere, whilst Case 9c is predominantly composed of CO$_2$. Whilst challenging from the ground, such observations are feasible in the near future by, for example, the James Webb Space Telescope (JWST). The Hubble Space Telescope (HST) is in principle sensitive to these signal levels, but lacks spectral coverage. The HST's WFC3 instrument's wavelength range, which is often used to observe exoplanets, is mostly only sensitive to H$_2$O and does not cover the major absorption bands of CO$_2$.
   \begin{figure*}
 \centering{
      \includegraphics[width=0.75\hsize]{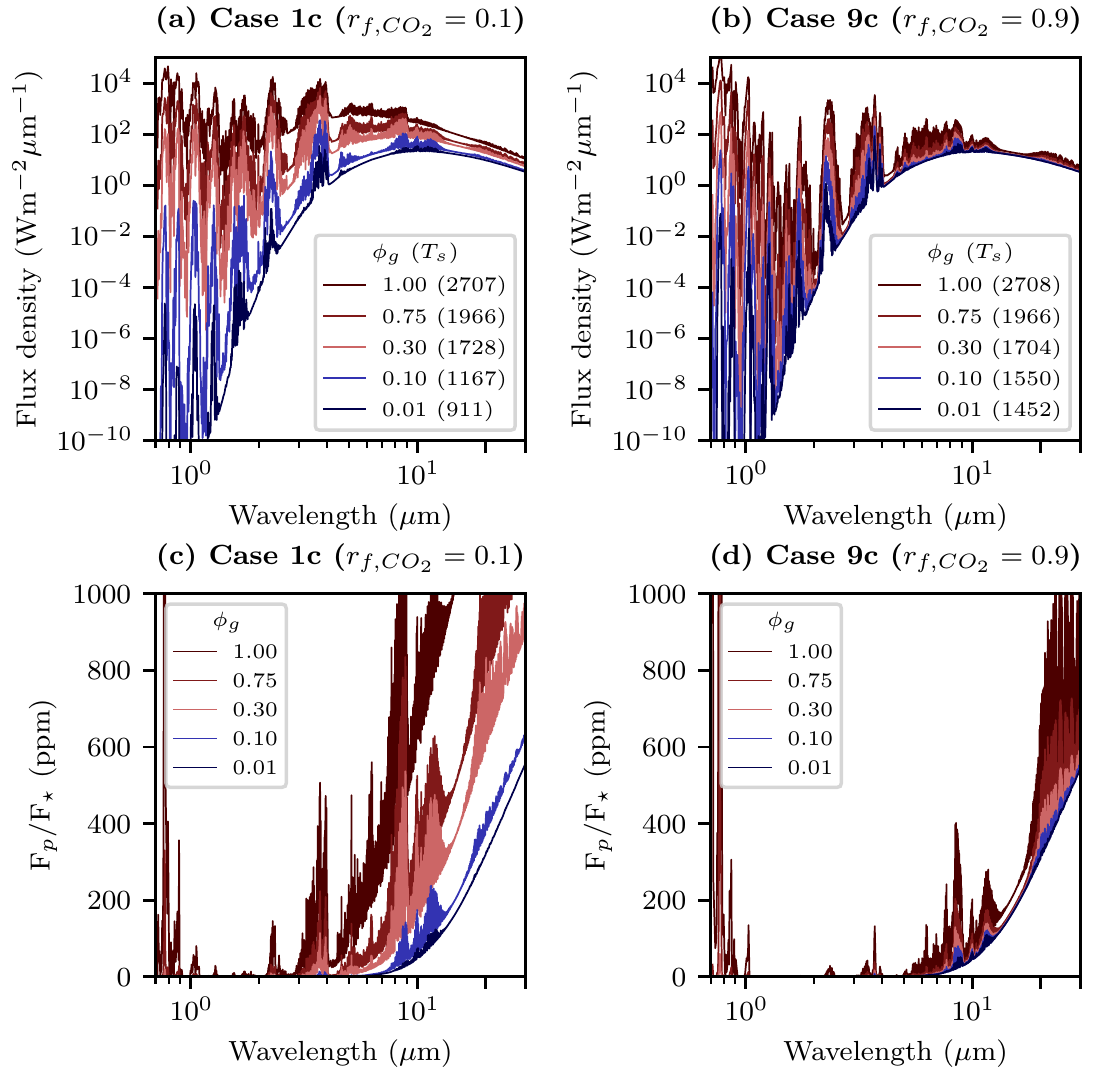}}
      \caption{(a, b) Prediction of emission spectra and (c, d) secondary eclipse signal for a planet orbiting an M-star at 1 AU with a mantle melt fraction $\phi_g=$1, 0.75, 0.3, 0.1, 0.01 and surface temperature $T_s$.  (a, c) Case 1c, (b, d) Case 9c.}
         \label{fig:emission}
   \end{figure*}

Figure~\ref{fig:emission} shows the corresponding emission spectra (upper panels) and secondary eclipse spectra (lower panels) for the same melt fractions as plotted for the transmission spectra.  We restrict the emission spectra to a planet orbiting at 1 AU around an M-star host.  Owing to the much higher luminosity of the G2V star, detecting the thermal emission spectrum of an orbiting planet would require a space-based interferometer or a space telescope with a starshade to mitigate the stellar light.  Whilst being theoretically investigated as a possible future space observatory (e.g. LUVOIR), no such mission is currently planned to be available within the foreseeable future.  The overall differences in the secondary eclipse spectra are of the order of several hundred ppm in the infrared wavelength range for both cases.  Such spectral variations are detectable in the near future with the JWST \citep{Gardner2006SSRv..123..485G}.  The different atmospheric compositions of the two cases can also be clearly distinguished; Case 1c features strong water absorption bands whilst Case 9c is dominated by the very deep CO$_2$ absorption band at 15 $\mu$m and has only muted water features.

Different melt fractions also show strong variations in the secondary eclipse spectra.  For a planet with a low melt fraction, the spectrum resembles a black body when compared to the spectrum for a planet with high melt fraction.  This is caused by the different temperature profiles.  For lower melt fractions and lower surface temperatures, the atmospheres become increasingly isothermal in the upper atmosphere (Fig.~\ref{fig:static}) which causes them to appear similar to a pure black-body spectrum.  Differences in the spectra can also be noticed at shorter wavelengths.  This part of the spectrum is determined by the back-scattering of the planet's host star radiation via molecular Rayleigh scattering rather than thermal emission.  Clear spectral signatures can only be seen for high melt fractions.  At low melt fractions, which is indicative of a significant outgassed atmosphere, the atmosphere is optically thick enough to absorb most of the stellar radiation which leads to a reduced back-scattering signal in the planetary spectrum.  Therefore, observing a planet in the infrared and at shorter wavelengths might enable us to differentiate observationally between the different atmospheric compositions and total surface pressures caused by the variations in melt fraction.

Combining both primary transit and secondary eclipse measurements together with mass estimates via radial velocity measurements can give constraints on the atmosphere and the planetary interior to a certain degree, at least for roughly Earth-sized planets orbiting small M-type host stars.  There are of course still degeneracies that can not be easily overcome, such as between the total surface pressure and the planetary surface radius for very massive atmospheres.  Transmission spectra inherently only provide information about the upper atmosphere and the overall planetary radius.  Emission spectra, meanwhile, are sensitive to the atmospheric temperature profile and pressure.  However, neither spectra can provide information about the very deep atmosphere if the planet possesses a massive atmosphere.  Thermal emission spectra originate from parts of the atmosphere where the optical depth integrated from the top of the atmosphere downwards is approximately unity.  Information from larger optical depths (and thus higher pressures) is therefore not accessible.
\subsubsection{Atmospheric escape}

We now evaluate thermal and non-thermal mechanisms of atmospheric escape in the context of our models.  In the thermal regime, we consider end members due to two heating mechanisms: solar heating at the surface (modified surface Jeans escape) and XUV heating of the upper atmosphere (energy-limited escape or hydrodynamic escape), both of which are discussed in \cite{JOY15}.  In the non-thermal regime we consider plasma-driven escape, where we focus on the observed mechanism of atmospheric sputtering \citep[e.g.][]{HWY81,J04,OJL19}.  Surface and upper atmospheric heating dominate light atmospheres (H, He), but nevertheless can be calculated for volatile species of any molar mass. In a test model (not shown), we implement escape due to surface heating \citep[Eq.~2b,][]{JOY15} and find that it has negligible influence on magma ocean cooling.  Even though the surface temperature is high, the molar mass of both H$_2$O and CO$_2$ is large which results in a large Jeans parameter and hence minimal Jeans escape in the absence of dissociation \citep[Eq.~1b,][]{JOY15}.  Upper atmospheric heating that is limited by XUV stellar irradiation is often used to approximate hydrodynamic escape at terrestrial exoplanets \citep[e.g.][]{BEW17}.  This may evaporate the secondary H$_2$O atmospheres that are produced from interior outgassing.  During the early Earth's transient magma ocean, the enhanced EUV irradiation can be more than one hundred times the present-day value, which results in around $10^{9}$g/s of H$_2$O loss \citep{CK17}.  This corresponds to roughly 1.5 kbar of water loss over 100 Myr, and may be the main driver of the desiccation of Venus.  Nevertheless, we caution that the physics of upper atmospheric heating is not well constrained for terrestrial planets and in particular the heating efficiency factor is necessarily assumed to be similar to large planets, although it may in fact be very different.  This is because the chemical speciation (e.g. CO$_2$, H$_2$, N$_2$) of terrestrial atmospheres regulates the heating of the upper atmosphere that is responsible for escape.

In-situ solar system observations reveal that plasma-driven heating is the dominant mechanism for atmospheric loss at all terrestrial planets at present day \citep[e.g.][]{ZC17,CK17}.  These data are particularly valuable to inform the loss rate of terrestrial atmospheres for planets orbiting M-dwarfs that are known to be strongly magnetic \citep{GDC17}.  MAVEN recently detected an oxygen escape rate of around $10^3$ g/s at Mars \citep{JBC18}, which when sputtered for 4.5 Gyr corresponds to a loss of approximately 8 bar of CO$_2$ \citep{LJZ92}.  For a planet at 1 AU, this yields a lower loss limit of approximately 18 bar of CO$_2$ due to sputtering alone.  Simulations that include plasma-driven losses as well as ion-molecular dissociation are therefore important for future long-term evolution models \citep[e.g.][]{EJB19,LBC19}.  In this regard, our model is an end-member that assumes retention of an outgassed atmosphere.  Importantly, atmospheric losses that occur over Gyr must be considered alongside geochemical cycling that also acts to deplete the atmospheric reservoir over geological timescales.  Consideration of both of these effects will be important for bridging short-term and long-term evolution models, and should form the basis of future coupled interior--atmosphere models.

\subsubsection{Condensation, clouds, and atmospheric convection}

We use a radiative-equilibrium solution with grey opacity to compute the atmospheric structure \citep{AM85}.  For an H$_2$O-dominated atmosphere, this solution is approximately valid for both the radiatively-equilibrated structure for small optical depths and the convectively-equilibrated structure for large optical depths \citep[Appendix A,][]{AM85}.  Although this formulation facilitates efficient calculations when integrated in an interior evolution model, it necessarily simplifies some aspects of the atmospheric structure.  For example, our atmosphere model does not consider condensation of H$_2$O, which can give rise to a moist troposphere that restricts the outgoing long-wavelength radiation (OLR) to around 300 Wm$^{-2}$ \citep[i.e. Nakajima's radiation limit,][]{NHA92}.  For an atmosphere with 300 bar of H$_2$O and 100 bar of CO$_2$, this limit is reached for surface temperatures less than around 1700 K for a cloud-free atmosphere and 2000 K if clouds are considered \citep{MSM17}.  In our models, the OLR always decreases monotonically with surface temperature, so exclusively incorporating a radiation limit would increase the cooling rate during the later stages of magma ocean crystallisation.  Conversely, clouds reduce the OLR by about a factor of 2 to 4, although their influence is diminished at high temperature \citep[Fig.~1,][]{MSM17}; hence, consideration of clouds alone would extend the lifetime of the magma ocean in our models.  It should also be noted that the interplay of radiation limits, solar insolation, and hydrodynamic escape, can produce an evolutionary dichotomy for initially molten planets that reside either side of a critical orbital distance \citep{HAG13}.

Clouds have a large impact on the spectra and surface temperature of Earth-like planets. They can scatter incident stellar radiation back to space but also contribute a greenhouse effect by trapping infrared radiation in the lower atmosphere \citep{Marley2013cctp.book..367M}.  The net radiative impact of clouds is determined by the size distribution of their particles, composition (for Earth-like planets: either solid or liquid water, or dry ice), optical depth, and also their location within the atmosphere.  In Earth's atmosphere, for example, low-level water clouds exhibit a net cooling effect on the surface, whilst high-level water ice clouds have a net greenhouse effect \citep{Kitzmann2010A&A...511A..66K}.  Determining the climatic impact of water clouds in the hot and massive atmospheres considered in this study requires knowledge of their formation mechanisms under strongly non-Earth-like atmospheric conditions.  Many of the clouds' properties depend on the spatial distribution of the initial cloud condensation nuclei, which are presently unknown for this class of atmosphere.  For example, it is not clear if condensation nuclei would be present in the region of the atmosphere where water can condense.  Since the vertical extension of the atmospheres in this study is large, nuclei would need to be transported over long distances from the surface to the high atmosphere (100s km versus less than 20 km for present-day Earth).  Therefore, due to the inherent complexity of treating cloud formation in hot and massive atmospheres, we choose to neglect them in this work.

Water clouds comparable to the ones found in Earth's atmosphere can mute spectral features of molecules and lower the overall OLR \citep{Kitzmann2011A&A...531A..62K}; a reduction in spectral flux is also reported for hot atmospheres with clouds above a magma ocean \citep[Figs.~8 and 10,][]{MSM17}.  However, water clouds can also enhance the spectral signals by increasing the amount of back-scattered light from the atmosphere to space \citep{Kitzmann2011A&A...534A..63K}.  Therefore, determining their actual impact again depends on their location in the atmosphere, composition, and particle size.  Clearly, clouds that form in parts of the atmosphere that are optically thick have a negligible impact on the thermal emission spectra.  As previously mentioned, since the details of cloud formation in the massive atmospheres that we consider are unknown, we do not take them into account for the calculation of the emission spectra (Fig. \ref{fig:emission}).  Besides their impact on the atmospheric emission and reflection, clouds are also able to influence the transmission spectra of planets.  Cloud layers high in the atmospheres of hot Jupiters and mini-Neptunes can increase the overall transit radius and mute the signatures of molecules in the transmission spectrum \citep{Marley2013cctp.book..367M}. For the atmospheres of terrestrial planets that we consider, clouds are usually found in the troposphere.  This region of the atmosphere is already mostly optically thick due to molecular absorption, such that the impact of clouds on the transmission spectra is negligible \citep{Kaltenegger2009ApJ...698..519K}.

We expect our primary result to remain robust despite the aforementioned complexities that will influence atmospheric structure.  To recap, the planetary radius of an Earth-like planet differs by 5\% depending whether it is molten or solid, and an outgassed atmosphere can introduce an additional radius perturbation of a few percent depending on the initial inventory of CO$_2$ and H$_2$O.  We observe the same physical mechanisms that control the atmospheric height as recently summarised in \cite{TEL19}: (1) total mass of the atmosphere, (2) temperature of the atmosphere, and (3) mean molar mass of the atmosphere (Fig.~\ref{fig:radius}).  The relatively late outgassing of H$_2$O compared to CO$_2$ is driven by the difference in solubility of the two volatiles in silicate melt.  This causes the planetary radius to expand for H$_2$O-dominated atmospheres by reducing the mean molar mass of the atmosphere, since contraction of the interior is largely complete at this stage.  In this regard, the atmospheric pressure--temperature profile is less important than the  atmospheric composition.  However, the pressure--temperature profile controls the condensation of H$_2$O and thereby the establishment of radiation limits, which coupled to volatile escape considerations could in principle produce different behaviour.  This is because preferential loss of lighter volatiles from the atmosphere will mitigate their ability to decrease the mean molar mass of the atmosphere, thereby reducing the expansion of the atmosphere.
\section{Conclusions}
We devise a coupled interior--atmosphere model to connect the evolution of a rocky planet to the observational nature of hot Earth-like planets with large outgassed atmospheres.  Our models are calibrated to give an outgassed atmosphere of 220 bar H$_2$O, and we vary the final CO$_2$ volume mixing ratio to explore the consequences of a CO$_2$ versus H$_2$O dominated atmosphere.  The relationship between the volatile mass and partial pressure is often misstated in the literature when more than a single volatile species is considered, so we provide clarification on the correct expression and calibrate previous results by running comparison cases.  It is anticipated that the correction will greatly affect atmospheres that contain species with strongly differing molar masses, and is perhaps magnified further if the species have similar solubilities.

We generate two suites of end-member models that produce different outgassing behaviour.  For early outgassing cases, typical of most models in the literature, outgassing is unimpeded by interior dynamics and magma ocean cooling generates a large outgassed atmosphere within a few million years.  For extended outgassing cases, the formation of a viscous lid at the surface prevents efficient outgassing, particularly of H$_2$O which has a higher solubility in silicate melt than CO$_2$.  This is because the cooling timescale is dictated by heat transport through the lid, and eventually the viscous flow of the solid mantle, rather than the radiative timescale of the atmosphere.  Including melt--solid separation can reduce the cooling timescale of extended outgassing cases by around an order of magnitude by mitigating the formation of the viscous lid.  This is probably a maximum reduction in the cooling timescale due to melt--solid separation, since we assume that melts are buoyant everywhere in the mantle and therefore substantial upwards draining of melt is facilitated.  For all cases, 90\% of the H$_2$O inventory remains in the interior and only readily outgasses once the global melt fraction drops below 10\%; this suggests that a large reservoir of H$_2$O can be retained in the interior until the very last stages of magma ocean crystallisation.  Therefore, the formation of a surface lid that restricts heat transport could control the outgassing rate of some volatiles from the interior of a rocky planet.

We combine the evolutionary models with static structure calculations to determine the radius of a rocky planet with varying degrees of melt from fully molten to fully solid and with an outgassed atmosphere.  Our results demonstrate that a hot molten planet can have a radius several percent larger ($\approx5\%$) than its equivalent solid counterpart, which may explain the larger radii of some close-in exoplanets.  Outgassing of a low molar mass species (such as H$_2$O, compared to CO$_2$) can combat the continual contraction of a planetary mantle and even marginally increase the planetary radius.  We further use our models to generate synthetic transmission and emission data to aid in the detection and characterisation of rocky planets via transits and secondary eclipses.  Atmospheres of terrestrial planets around M-stars that are dominated by CO$_2$ or H$_2$O can be distinguished by future observing facilities that have extended wavelength coverage (e.g. JWST).  Incomplete magma ocean crystallisation and full or part retention of an early outgassed atmosphere---as may be the case for close-in terrestrial planets---should be considered in the interpretation of observational data.

\begin{acknowledgements}
DJB acknowledges Swiss National Science Foundation (SNSF) Ambizione Grant 173992.  ASW acknowledges the Turner postdoctoral fellowship and OAC-15-50482 in supporting this work.  PS acknowledges financial support from the Swiss University Conference and the Swiss Council of Federal Institutes of Technology through the Platform for Advanced Scientific Computing (PASC) programme.  CD acknowledges SNSF Ambizione Grant PZ00P2\_174028.  AVO acknowledges support from the SNSF grant BSSGI0$\_$155816 ``PlanetsInTime''.  This research was partly inspired by discussions and interactions within the framework of the National Center for Competence in Research (NCCR) PlanetS supported by the SNSF.  The calculations were performed on UBELIX (http://www.id.unibe.ch/hpc), the HPC cluster at the University of Bern.  We thank T. Lichtenberg for comments on an earlier draft of the manuscript and B.-O. Demory for discussions.  Constructive critique from an anonymous reviewer further enhanced the manuscript.
\end{acknowledgements}

%
%

\bibliographystyle{aa}
\bibliography{refs}

\begin{thebibliography}{81}
\expandafter\ifx\csname natexlab\endcsname\relax\def\natexlab#1{#1}\fi

\bibitem[{Abe(1993)}]{ABE93}
Abe, Y. 1993, in Evolution of the Earth and Planets, ed. E.~Takahashi,
  R.~Jeanloz, \& D.~Rubie, Vol.~74 (AGU, {W}ashington, {D.C.}), 41--54

\bibitem[{Abe(1995)}]{ABE95}
Abe, Y. 1995, in The Earth's Central Part: Its Structure and Dynamics, ed.
  T.~Yukutake (Terra Sci. Pub. Com., Tokyo), 215--230

\bibitem[{Abe(1997)}]{ABE97}
Abe, Y. 1997, Phys. Earth Planet. Inter., 100, 27

\bibitem[{Abe \& Matsui(1985)}]{AM85}
Abe, Y. \& Matsui, T. 1985, J. Geophys. Res.-So. Ea., 90, C545

\bibitem[{Andrault {et~al.}(2011)Andrault, Bolfan-Casanova, Nigro, Bouhifd,
  Garbarino, \& Mezouar}]{ABL11}
Andrault, D., Bolfan-Casanova, N., Nigro, G.~L., {et~al.} 2011, Earth Planet.
  Sci. Lett., 304, 251

\bibitem[{Ballmer {et~al.}(2017)Ballmer, Louren\c{c}o, Hirose, Caracas, \&
  Nomura}]{BLH17}
Ballmer, M.~D., Louren\c{c}o, D.~L., Hirose, K., Caracas, R., \& Nomura, R.
  2017, Geochem. Geophy. Geosys., 18, 2785

\bibitem[{{Barber} {et~al.}(2006){Barber}, {Tennyson}, {Harris}, \&
  {Tolchenov}}]{Barber2006MNRAS.368.1087B}
{Barber}, R.~J., {Tennyson}, J., {Harris}, G.~J., \& {Tolchenov}, R.~N. 2006,
  MNRAS, 368, 1087

\bibitem[{{Barclay} {et~al.}(2018){Barclay}, {Pepper}, \& {Quintana}}]{BPQ18}
{Barclay}, T., {Pepper}, J., \& {Quintana}, E.~V. 2018, ApJS, 239, 2

\bibitem[{Bolfan-Casanova {et~al.}(2003)Bolfan-Casanova, Keppler, \&
  Rubie}]{BKR03}
Bolfan-Casanova, N., Keppler, H., \& Rubie, D.~C. 2003, Geophys. Res. Lett., 30

\bibitem[{Bolmont {et~al.}(2013)Bolmont, Selsis, Raymond, Leconte, Hersant,
  Maurin, \& Pericaud}]{BSR13}
Bolmont, E., Selsis, F., Raymond, S.~N., {et~al.} 2013, A\&A, 556, A17

\bibitem[{Bonati {et~al.}(2019)Bonati, Lichtenberg, Bower, Timpe, \&
  Quanz}]{BLB19}
Bonati, I., Lichtenberg, T., Bower, D.~J., Timpe, M., \& Quanz, S. 2019, A\&A,
  621

\bibitem[{Bond {et~al.}(2010)Bond, O'Brien, \& Lauretta}]{BOL10}
Bond, J.~C., O'Brien, D.~P., \& Lauretta, D.~S. 2010, ApJ, 715, 1050

\bibitem[{Bourrier {et~al.}(2017)Bourrier, Ehrenreich, Wheatley, Bolmont,
  Gillon, de~Wit, Burgasser, Jehin, Queloz, \& Triaud}]{BEW17}
Bourrier, V., Ehrenreich, D., Wheatley, P.~J., {et~al.} 2017, A\&A, 599, L3

\bibitem[{Bower {et~al.}(2018)Bower, Sanan, \& Wolf}]{BSW18}
Bower, D.~J., Sanan, P., \& Wolf, A.~S. 2018, Phys. Earth Planet. Inter., 274,
  49

\bibitem[{Carroll \& Holloway(1994)}]{CHM94}
Carroll, M.~R. \& Holloway, J.~R., eds. 1994, Volatiles in Magmas, Vol.~30
  (Mineralogical Society of America)

\bibitem[{Catling \& Kasting(2017)}]{CK17}
Catling, D.~C. \& Kasting, J.~F. 2017, Atmospheric Evolution on Inhabited and
  Lifeless Worlds (Cambridge University Press)

\bibitem[{Dorn {et~al.}(2018)Dorn, Bower, \& Rozel}]{DBR18}
Dorn, C., Bower, D.~J., \& Rozel, A. 2018, in Handbook of Exoplanets, ed. H.~J.
  Deeg \& J.~A. Belmonte (Springer International Publishing)

\bibitem[{Dorn {et~al.}(2019)Dorn, Harrison, Bonsor, \& Hands}]{DHB19}
Dorn, C., Harrison, J., Bonsor, A., \& Hands, T. 2019, MNRAS, 484, 712

\bibitem[{Dorn {et~al.}(2016)Dorn, Hinkel, \& Venturini}]{DHV16}
Dorn, C., Hinkel, N.~R., \& Venturini, J. 2016, A\&A, 597, A38

\bibitem[{Egan {et~al.}(2019)Egan, Jarvinen, \& Brain}]{EJB19}
Egan, H., Jarvinen, R., \& Brain, D. 2019, MNRAS, 486, 1283

\bibitem[{Elkins-Tanton(2008)}]{ET08}
Elkins-Tanton, L. 2008, Earth Planet. Sci. Lett., 271, 181

\bibitem[{Elkins-Tanton(2012)}]{E12}
Elkins-Tanton, L.~T. 2012, Ann. Rev. Earth Planet. Sci., 40, 113

\bibitem[{Fiquet {et~al.}(2010)Fiquet, Auzende, Siebert, Corgne, Bureau, Ozawa,
  \& Garbarino}]{FAS10}
Fiquet, G., Auzende, A.~L., Siebert, J., {et~al.} 2010, Sci., 329, 1516

\bibitem[{{Gardner} {et~al.}(2006){Gardner}, {Mather}, {Clampin}, {Doyon},
  {Greenhouse}, {Hammel}, {Hutchings}, {Jakobsen}, {Lilly}, {Long}, {Lunine},
  {McCaughrean}, {Mountain}, {Nella}, {Rieke}, {Rieke}, {Rix}, {Smith},
  {Sonneborn}, {Stiavelli}, {Stockman}, {Windhorst}, \&
  {Wright}}]{Gardner2006SSRv..123..485G}
{Gardner}, J.~P., {Mather}, J.~C., {Clampin}, M., {et~al.} 2006, \ssr, 123, 485

\bibitem[{Garraffo {et~al.}(2017)Garraffo, Drake, Cohen, Alvarado-G{\'{o}}mez,
  \& Moschou}]{GDC17}
Garraffo, C., Drake, J.~J., Cohen, O., Alvarado-G{\'{o}}mez, J.~D., \& Moschou,
  S.~P. 2017, ApJ, 843, L33

\bibitem[{{Grimm} \& {Heng}(2015)}]{Grimm2015ApJ...808..182G}
{Grimm}, S.~L. \& {Heng}, K. 2015, ApJ, 808, 182

\bibitem[{Haff {et~al.}(1981)Haff, Watson, \& Yung}]{HWY81}
Haff, P.~K., Watson, C.~C., \& Yung, Y.~L. 1981, J. Geophys. Res.-Space, 86,
  6933

\bibitem[{Hamano {et~al.}(2013)Hamano, Abe, \& Genda}]{HAG13}
Hamano, K., Abe, Y., \& Genda, H. 2013, Nat., 497, 607

\bibitem[{{Hamre} {et~al.}(2013){Hamre}, {Stamnes}, {Stamnes}, \&
  {Stamnes}}]{Hamre2013AIPC.1531..923H}
{Hamre}, B., {Stamnes}, S., {Stamnes}, K., \& {Stamnes}, J.~J. 2013, in
  American Institute of Physics Conference Series, Vol. 1531, American
  Institute of Physics Conference Series, 923--926

\bibitem[{{Heng} \& {Kitzmann}(2017)}]{Heng2017MNRAS.470.2972H}
{Heng}, K. \& {Kitzmann}, D. 2017, Mon. Not. R. Astron. Soc., 470, 2972

\bibitem[{Hirschmann(2000)}]{H00}
Hirschmann, M.~M. 2000, Geochem. Geophy. Geosys., 1

\bibitem[{Jakosky {et~al.}(2018)Jakosky, Brain, Chaffin, Curry, Deighan,
  Grebowsky, Halekas, Leblanc, Lillis, Luhmann, Andersson, Andre, Andrews,
  Baird, Baker, Bell, Benna, Bhattacharyya, Bougher, Bowers, Chamberlin,
  Chaufray, Clarke, Collinson, Combi, Connerney, Connour, Correira, Crabb,
  Crary, Cravens, Crismani, Delory, Dewey, DiBraccio, Dong, Dong, Dunn, Egan,
  Elrod, England, Eparvier, Ergun, Eriksson, Esman, Espley, Evans, Fallows,
  Fang, Fillingim, Flynn, Fogle, Fowler, Fox, Fujimoto, Garnier, Girazian,
  Groeller, Gruesbeck, Hamil, Hanley, Hara, Harada, Hermann, Holmberg,
  Holsclaw, Houston, Inui, Jain, Jolitz, Kotova, Kuroda, Larson, Lee, Lee,
  Lefevre, Lentz, Lo, Lugo, Ma, Mahaffy, Marquette, Matsumoto, Mayyasi,
  Mazelle, McClintock, McFadden, Medvedev, Mendillo, Meziane, Milby, Mitchell,
  Modolo, Montmessin, Nagy, Nakagawa, Narvaez, Olsen, Pawlowski, Peterson,
  Rahmati, Roeten, Romanelli, Ruhunusiri, Russell, Sakai, Schneider, Seki,
  Sharrar, Shaver, Siskind, Slipski, Soobiah, Steckiewicz, Stevens, Stewart,
  Stiepen, Stone, Tenishev, Terada, Terada, Thiemann, Tolson, Toth, Trovato,
  Vogt, Weber, Withers, Xu, Yelle, Yiğit, \& Zurek}]{JBC18}
Jakosky, B., Brain, D., Chaffin, M., {et~al.} 2018, Icarus, 315, 146

\bibitem[{Johnson(2004)}]{J04}
Johnson, R.~E. 2004, ApJ, 609, L99

\bibitem[{Johnson {et~al.}(2015)Johnson, Oza, Young, Volkov, \&
  Schmidt}]{JOY15}
Johnson, R.~E., Oza, A., Young, L.~A., Volkov, A.~N., \& Schmidt, C. 2015, ApJ,
  809, 43

\bibitem[{{Kaltenegger} \& {Traub}(2009)}]{Kaltenegger2009ApJ...698..519K}
{Kaltenegger}, L. \& {Traub}, W.~A. 2009, ApJ, 698, 519

\bibitem[{Kislyakova {et~al.}(2017)Kislyakova, Noack, Johnstone, Zaitsev,
  Fossati, Lammer, Khodachenko, Odert, \& Güdel}]{KNJ17}
Kislyakova, K.~G., Noack, L., Johnstone, C.~P., {et~al.} 2017, Nat. Astron., 1,
  878

\bibitem[{Kite {et~al.}(2016)Kite, Fegley~Jr, Schaefer, \&
  Gaidos}]{kite2016atmosphere}
Kite, E.~S., Fegley~Jr, B., Schaefer, L., \& Gaidos, E. 2016, ApJ, 828, 80

\bibitem[{{Kitzmann}(2017)}]{Kitzmann2017A&A...600A.111K}
{Kitzmann}, D. 2017, \aap, 600, A111

\bibitem[{{Kitzmann} {et~al.}(2011{\natexlab{a}}){Kitzmann}, {Patzer}, {von
  Paris}, {Godolt}, \& {Rauer}}]{Kitzmann2011A&A...531A..62K}
{Kitzmann}, D., {Patzer}, A.~B.~C., {von Paris}, P., {Godolt}, M., \& {Rauer},
  H. 2011{\natexlab{a}}, A\&A, 531, A62

\bibitem[{{Kitzmann} {et~al.}(2011{\natexlab{b}}){Kitzmann}, {Patzer}, {von
  Paris}, {Godolt}, \& {Rauer}}]{Kitzmann2011A&A...534A..63K}
{Kitzmann}, D., {Patzer}, A.~B.~C., {von Paris}, P., {Godolt}, M., \& {Rauer},
  H. 2011{\natexlab{b}}, A\&A, 534, A63

\bibitem[{{Kitzmann} {et~al.}(2010){Kitzmann}, {Patzer}, {von Paris}, {Godolt},
  {Stracke}, {Gebauer}, {Grenfell}, \& {Rauer}}]{Kitzmann2010A&A...511A..66K}
{Kitzmann}, D., {Patzer}, A.~B.~C., {von Paris}, P., {et~al.} 2010, A\&A, 511,
  A66

\bibitem[{Korenaga(2010)}]{KJUN10}
Korenaga, J. 2010, ApJL, 725, L43

\bibitem[{Labrosse {et~al.}(2007)Labrosse, Hernlund, \& Coltice}]{LHC07}
Labrosse, S., Hernlund, J.~W., \& Coltice, N. 2007, Nat., 450, 866

\bibitem[{Leblanc {et~al.}(2019)Leblanc, Benna, Chaufray, Martinez, Lillis,
  Curry, Elrod, Mahaffy, Modolo, Luhmann, \& Jakosky}]{LBC19}
Leblanc, F., Benna, M., Chaufray, J.~Y., {et~al.} 2019, Geophys. Res. Lett.,
  46, 4144

\bibitem[{Lebrun {et~al.}(2013)Lebrun, Massol, Chassefi\`{e}re, Davaille,
  Marcq, Sarda, Leblanc, \& Brandeis}]{LMC13}
Lebrun, T., Massol, H., Chassefi\`{e}re, E., {et~al.} 2013, J. Geophys.
  Res.-Planet., 118, 1155

\bibitem[{Lichtenberg {et~al.}(2016)Lichtenberg, Golabek, Gerya, \&
  Meyer}]{LGG16}
Lichtenberg, T., Golabek, G.~J., Gerya, T.~V., \& Meyer, M.~R. 2016, Icarus,
  274, 350

\bibitem[{Luhmann {et~al.}(1992)Luhmann, Johnson, \& Zhang}]{LJZ92}
Luhmann, J.~G., Johnson, R.~E., \& Zhang, M. H.~G. 1992, Geophys. Res. Lett.,
  19, 2151

\bibitem[{Lupu {et~al.}(2014)Lupu, Zahnle, Marley, Schaefer, Fegley, Morley,
  Cahoy, Freedman, \& Fortney}]{LZM14}
Lupu, R.~E., Zahnle, K., Marley, M.~S., {et~al.} 2014, ApJ, 784, 27

\bibitem[{Marcq {et~al.}(2017)Marcq, Salvador, Massol, \& Davaille}]{MSM17}
Marcq, E., Salvador, A., Massol, H., \& Davaille, A. 2017, J. Geophys.
  Res.-Planet., 122, 1539

\bibitem[{{Marley} {et~al.}(2013){Marley}, {Ackerman}, {Cuzzi}, \&
  {Kitzmann}}]{Marley2013cctp.book..367M}
{Marley}, M.~S., {Ackerman}, A.~S., {Cuzzi}, J.~N., \& {Kitzmann}, D. 2013, in
  Comparative Climatology of Terrestrial Planets, ed. S.~J. {Mackwell}, A.~A.
  {Simon-Miller}, J.~W. {Harder}, \& M.~A. {Bullock}, Space Science Series
  (University of Arizona Press), 367--391

\bibitem[{Maurice {et~al.}(2017)Maurice, Tosi, Samuel, Plesa, H\"{u}ttig, \&
  Breuer}]{MTS17}
Maurice, M., Tosi, N., Samuel, H., {et~al.} 2017, J. Geophys. Res.-Planet.,
  122, 577

\bibitem[{Miller-Ricci {et~al.}(2009)Miller-Ricci, Meyer, Seager, \&
  Elkins-Tanton}]{MMS09}
Miller-Ricci, E., Meyer, M.~R., Seager, S., \& Elkins-Tanton, L. 2009, ApJ,
  704, 770

\bibitem[{Monteux {et~al.}(2016)Monteux, Andrault, \& Samuel}]{MAS16}
Monteux, J., Andrault, D., \& Samuel, H. 2016, Earth Planet. Sci. Lett., 448,
  140

\bibitem[{Mosenfelder {et~al.}(2009)Mosenfelder, Asimow, Frost, Rubie, \&
  Ahrens}]{M09}
Mosenfelder, J.~L., Asimow, P.~D., Frost, D.~J., Rubie, D.~C., \& Ahrens, T.~J.
  2009, J. Geophys. Res.-So. Ea., 114

\bibitem[{Nakajima \& Stevenson(2015)}]{NS15}
Nakajima, M. \& Stevenson, D.~J. 2015, Earth Planet. Sci. Lett., 427, 286

\bibitem[{Nakajima {et~al.}(1992)Nakajima, Hayashi, \& Abe}]{NHA92}
Nakajima, S., Hayashi, Y.-Y., \& Abe, Y. 1992, J. Atmospheric Sci., 49, 2256

\bibitem[{Nikolaou {et~al.}(2019)Nikolaou, Katyal, Tosi, Godolt, Grenfell, \&
  Rauer}]{NKT19}
Nikolaou, A., Katyal, N., Tosi, N., {et~al.} 2019, ApJ, 875, 11

\bibitem[{Oza {et~al.}(2019)Oza, Johnson, Lellouch, Schmidt, Schneider, Huang,
  Gamborino, Gebek, Wyttenbach, Demory, Mordasini, Saxena, Dubois, Moullet, \&
  Thomas}]{OJL19}
Oza, A.~V., Johnson, R.~E., Lellouch, E., {et~al.} 2019, ApJ, accepted

\bibitem[{Pan {et~al.}(1991)Pan, Holloway, \& Hervig}]{PHH91}
Pan, V., Holloway, J.~R., \& Hervig, R.~L. 1991, Geochim. Cosmochim. Ac., 55,
  1587

\bibitem[{Pierrehumbert(2011)}]{P11}
Pierrehumbert, R.~T. 2011, Principles of Planetary Climate (Cambridge
  University Press)

\bibitem[{Pujol \& North(2003)}]{PN03}
Pujol, T. \& North, G.~R. 2003, Tellus A, 55, 328

\bibitem[{{Rauer} {et~al.}(2014){Rauer}, {Catala}, {Aerts}, {Appourchaux},
  {Benz}, {Brandeker}, {Christensen-Dalsgaard}, {Deleuil}, {Gizon}, {Goupil},
  {G{\"u}del}, {Janot-Pacheco}, {Mas-Hesse}, {Pagano}, {Piotto}, {Pollacco},
  {Santos}, {Smith}, {Su{\'a}rez}, {Szab{\'o}}, {Udry}, {Adibekyan}, {Alibert},
  {Almenara}, {Amaro-Seoane}, {Eiff}, {Asplund}, {Antonello}, {Barnes},
  {Baudin}, {Belkacem}, {Bergemann}, {Bihain}, {Birch}, {Bonfils}, {Boisse},
  {Bonomo}, {Borsa}, {Brand{\~a}o}, {Brocato}, {Brun}, {Burleigh}, {Burston},
  {Cabrera}, {Cassisi}, {Chaplin}, {Charpinet}, {Chiappini}, {Church},
  {Csizmadia}, {Cunha}, {Damasso}, {Davies}, {Deeg}, {D{\'{\i}}az}, {Dreizler},
  {Dreyer}, {Eggenberger}, {Ehrenreich}, {Eigm{\"u}ller}, {Erikson}, {Farmer},
  {Feltzing}, {de Oliveira Fialho}, {Figueira}, {Forveille}, {Fridlund},
  {Garc{\'{\i}}a}, {Giommi}, {Giuffrida}, {Godolt}, {Gomes da Silva},
  {Granzer}, {Grenfell}, {Grotsch-Noels}, {G{\"u}nther}, {Haswell}, {Hatzes},
  {H{\'e}brard}, {Hekker}, {Helled}, {Heng}, {Jenkins}, {Johansen},
  {Khodachenko}, {Kislyakova}, {Kley}, {Kolb}, {Krivova}, {Kupka}, {Lammer},
  {Lanza}, {Lebreton}, {Magrin}, {Marcos-Arenal}, {Marrese}, {Marques},
  {Martins}, {Mathis}, {Mathur}, {Messina}, {Miglio}, {Montalban}, {Montalto},
  {Monteiro}, {Moradi}, {Moravveji}, {Mordasini}, {Morel}, {Mortier},
  {Nascimbeni}, {Nelson}, {Nielsen}, {Noack}, {Norton}, {Ofir}, {Oshagh},
  {Ouazzani}, {P{\'a}pics}, {Parro}, {Petit}, {Plez}, {Poretti}, {Quirrenbach},
  {Ragazzoni}, {Raimondo}, {Rainer}, {Reese}, {Redmer}, {Reffert},
  {Rojas-Ayala}, {Roxburgh}, {Salmon}, {Santerne}, {Schneider}, {Schou},
  {Schuh}, {Schunker}, {Silva-Valio}, {Silvotti}, {Skillen}, {Snellen}, {Sohl},
  {Sousa}, {Sozzetti}, {Stello}, {Strassmeier}, {{\v S}vanda}, {Szab{\'o}},
  {Tkachenko}, {Valencia}, {Van Grootel}, {Vauclair}, {Ventura}, {Wagner},
  {Walton}, {Weingrill}, {Werner}, {Wheatley}, \& {Zwintz}}]{RCA14}
{Rauer}, H., {Catala}, C., {Aerts}, C., {et~al.} 2014, Exp. Astron., 38, 249

\bibitem[{{Rothman} {et~al.}(2010){Rothman}, {Gordon}, {Barber}, {Dothe},
  {Gamache}, {Goldman}, {Perevalov}, {Tashkun}, \&
  {Tennyson}}]{Rothman2010JQSRT.111.2139R}
{Rothman}, L.~S., {Gordon}, I.~E., {Barber}, R.~J., {et~al.} 2010, J. Quant.
  Spectr. Rad. Transf., 111, 2139

\bibitem[{Ruedas(2017)}]{RUE17}
Ruedas, T. 2017, Geochem. Geophy. Geosys., 18, 3530

\bibitem[{Salvador {et~al.}(2017)Salvador, Massol, Davaille, Marcq, Sarda, \&
  Chassefi\`{e}re}]{SMD17}
Salvador, A., Massol, H., Davaille, A., {et~al.} 2017, J. Geophys.
  Res.-Planet., 122, 1458

\bibitem[{Schaefer \& Elkins-Tanton(2018)}]{SE18}
Schaefer, L. \& Elkins-Tanton, L.~T. 2018, Phil. Trans. R. Soc. Lond., Ser. A,
  376, 20180109

\bibitem[{Schaefer \& Fegley(2010)}]{SF10}
Schaefer, L. \& Fegley, B. 2010, Icarus, 208, 438

\bibitem[{Schlichting \& Mukhopadhyay(2018)}]{SM18}
Schlichting, H.~E. \& Mukhopadhyay, S. 2018, Space Sci. Rev., 214, 34

\bibitem[{Shcheka {et~al.}(2006)Shcheka, Wiedenbeck, Frost, \& Keppler}]{SWF06}
Shcheka, S.~S., Wiedenbeck, M., Frost, D.~J., \& Keppler, H. 2006, Earth
  Planet. Sci. Lett., 245, 730

\bibitem[{Solomatov(2000)}]{VSS00}
Solomatov, V.~S. 2000, in Origin of the Earth and Moon, ed. R.~M. Canup \&
  K.~Righter, Space Science Series (University of Arizona Press), 323--338

\bibitem[{Solomatov(2007)}]{SOLO07}
Solomatov, V.~S. 2007, in Treatise on Geophysics, ed. G.~Schubert, Vol. 9:
  Evolution of the Earth (Elsevier), 91--119

\bibitem[{Stixrude {et~al.}(2009)Stixrude, de~Koker, Sun, Mookherjee, \&
  Karki}]{SKS09}
Stixrude, L., de~Koker, N., Sun, N., Mookherjee, M., \& Karki, B.~B. 2009,
  Earth Planet. Sci. Lett., 278, 226

\bibitem[{Stothers \& Chin(1997)}]{SC97}
Stothers, R.~B. \& Chin, C. 1997, ApJL, 478, L103

\bibitem[{Tinetti {et~al.}(2018)Tinetti, Drossart, Eccleston, Hartogh, Heske,
  Leconte, Micela, Ollivier, Pilbratt, Puig, Turrini, Vandenbussche,
  Wolkenberg, Beaulieu, Buchave, Ferus, Griffin, Guedel, Justtanont, Lagage,
  Machado, Malaguti, Min, N{\o}rgaard-Nielsen, Rataj, Ray, Ribas, Swain, Szabo,
  Werner, Barstow, Burleigh, Cho, du~Foresto, Coustenis, Decin, Encrenaz,
  Galand, Gillon, Helled, Morales, Mu{\~{n}}oz, Moneti, Pagano, Pascale,
  Piccioni, Pinfield, Sarkar, Selsis, Tennyson, Triaud, Venot, Waldmann,
  Waltham, Wright, Amiaux, Augu{\`e}res, Berth{\'e}, Bezawada, Bishop, Bowles,
  Coffey, Colom{\'e}, Crook, Crouzet, Da~Peppo, Sanz, Focardi, Frericks, Hunt,
  Kohley, Middleton, Morgante, Ottensamer, Pace, Pearson, Stamper, Symonds,
  Rengel, Renotte, Ade, Affer, Alard, Allard, Altieri, Andr{\'e}, Arena,
  Argyriou, Aylward, Baccani, Bakos, Banaszkiewicz, Barlow, Batista, Bellucci,
  Benatti, Bernardi, B{\'e}zard, Blecka, Bolmont, Bonfond, Bonito, Bonomo,
  Brucato, Brun, Bryson, Bujwan, Casewell, Charnay, Pestellini, Chen,
  Ciaravella, Claudi, Cl{\'e}dassou, Damasso, Damiano, Danielski, Deroo,
  Di~Giorgio, Dominik, Doublier, Doyle, Doyon, Drummond, Duong, Eales, Edwards,
  Farina, Flaccomio, Fletcher, Forget, Fossey, Fr{\"a}nz, Fujii,
  Garc{\'i}a-Piquer, Gear, Geoffray, G{\'e}rard, Gesa, Gomez, Graczyk,
  Griffith, Grodent, Guarcello, Gustin, Hamano, Hargrave, Hello, Heng, Herrero,
  Hornstrup, Hubert, Ida, Ikoma, Iro, Irwin, Jarchow, Jaubert, Jones, Julien,
  Kameda, Kerschbaum, Kervella, Koskinen, Krijger, Krupp, Lafarga, Landini,
  Lellouch, Leto, Luntzer, Rank-L{\"u}ftinger, Maggio, Maldonado, Maillard,
  Mall, Marquette, Mathis, Maxted, Matsuo, Medvedev, Miguel, Minier, Morello,
  Mura, Narita, Nascimbeni, Nguyen~Tong, Noce, Oliva, Palle, Palmer, Pancrazzi,
  Papageorgiou, Parmentier, Perger, Petralia, Pezzuto, Pierrehumbert,
  Pillitteri, Piotto, Pisano, Prisinzano, Radioti, R{\'e}ess, Rezac, Rocchetto,
  Rosich, Sanna, Santerne, Savini, Scandariato, Sicardy, Sierra, Sindoni, Skup,
  Snellen, Sobiecki, Soret, Sozzetti, Stiepen, Strugarek, Taylor, Taylor,
  Terenzi, Tessenyi, Tsiaras, Tucker, Valencia, Vasisht, Vazan, Vilardell,
  Vinatier, Viti, Waters, Wawer, Wawrzaszek, Whitworth, Yung, Yurchenko,
  Osorio, Zellem, Zingales, \& Zwart}]{TDE18}
Tinetti, G., Drossart, P., Eccleston, P., {et~al.} 2018, Exp. Astron., 46, 135

\bibitem[{{Turbet} {et~al.}(2019){Turbet}, {Ehrenreich}, {Lovis}, {Bolmont}, \&
  {Fauchez}}]{TEL19}
{Turbet}, M., {Ehrenreich}, D., {Lovis}, C., {Bolmont}, E., \& {Fauchez}, T.
  2019, arXiv e-prints, arXiv:1906.03527

\bibitem[{Turcotte \& Schubert(2014)}]{TS14}
Turcotte, D. \& Schubert, G. 2014, Geodynamics (Cambridge University Press)

\bibitem[{Valencia {et~al.}(2007)Valencia, Sasselov, \& O'Connell}]{VSO07}
Valencia, D., Sasselov, D.~D., \& O'Connell, R.~J. 2007, ApJ, 656, 545

\bibitem[{Wicks {et~al.}(2018)Wicks, Smith, Fratanduono, Coppari, Kraus,
  Newman, Rygg, Eggert, \& Duffy}]{WSF18}
Wicks, J.~K., Smith, R.~F., Fratanduono, D.~E., {et~al.} 2018, Sci. Adv., 4

\bibitem[{Wolf \& Bower(2018)}]{WB18}
Wolf, A.~S. \& Bower, D.~J. 2018, Phys. Earth Planet. Inter., 278, 59

\bibitem[{Yamamoto \& Onishi(1952)}]{YO52}
Yamamoto, G. \& Onishi, G. 1952, J. Meteor., 9, 415

\bibitem[{Zahnle \& Catling(2017)}]{ZC17}
Zahnle, K.~J. \& Catling, D.~C. 2017, ApJ, 843, 122

\end{thebibliography}






   
  



\begin{appendix}
\section{Volatile mass balance and evolution}
\label{app:volmass}
The volatile mass balance is expressed following \cite{LMC13}, except we correct the expression for the relationship between the partial pressure of a volatile and its mass in the atmosphere (Eq.~\ref{eq:pmass}).  For a given volatile $v$,
\begin{equation}
X_v (k_v M^s + M^l) + 4 \pi R_p^2 \left( \frac{\mu_v}{\bar{\mu}} \right) \frac{p_v}{g} = X_v^0 M^m\,,
\label{eq:volmass}
\end{equation}
where $X_v$ is the volatile concentration in the melt (liquid) phase, $k_v$ distribution coefficient between solid and melt, $M^s$ mantle mass of solid, $M^l$ mantle mass of melt, $R_p$ planetary (surface) radius, $g$ surface gravity, $\mu_v$ volatile molar mass, $\bar{\mu}$ mean molar mass of the atmosphere, $p_v$ (surface) partial pressure of the volatile, $X_v^0$ initial volatile concentration in the mantle, and $M^m = M^s + M^l$ is total mantle mass.  The total surface pressure $P_s$ is given by Dalton's law, which for $n$ volatiles is
\begin{equation}
P_s = \sum_v^n p_v\,.
\end{equation}
The mean molar mass of the atmosphere is
\begin{equation}
\bar{\mu} = \sum_v^n \left( \frac{p_v\ \mu_v}{P_s} \right)\,.
\end{equation}
For an atmosphere with $n$ volatiles, the time derivative of Eq.~\ref{eq:volmass} is
\begin{align}
&\frac{d X_v}{d t} \left(k_v M^m + (1-k_v) M^l \right)
+ X_v (1-k_v) \frac{d M^l}{d t} \nonumber \\
& + \frac{4 \pi R_p^2 \mu_v}{g}
\left[\frac{1}{\bar{\mu}} \frac{d p_v}{d t}
-\frac{p_v}{\bar{\mu}^2} \sum_w^n \frac{\mu_w}{P_s} \left( \frac{d p_w}{d t} - \frac{p_w}{P_s} \frac{d P_s}{d t} \right)
\right] = 0\,.
\label{eq:volmassdt}
\end{align}
Time-derivatives of $p_v$, and by extension $P_s$, can be reduced to time-derivatives of $X_v$ by applying the chain rule to Eq.~\ref{eq:partial}.  To show the origin of the discrepancy between the formulation of previous studies and this study, we evaluate Eq.~\ref{eq:volmassdt} for $n=2$ volatiles.  For notional clarity, we now adopt a subscript of $H$ to denote H$_2$O and $C$ to denote CO$_2$, although in actuality the subscripts could refer to any two arbitrary volatile species.  Eq.~\ref{eq:volmassdt} now becomes
\begin{align}
&\frac{d X_v}{d t} \left(k_v M^m + (1-k_v) M^l \right)
+ X_v (1-k_v) \frac{d M^l}{d t} \nonumber \\
& + \frac{4 \pi R_p^2 \mu_v}{g}
\left[\frac{1}{\bar{\mu}} \frac{d p_v}{d t}
-\frac{p_v}{\bar{\mu}^2} \frac{(\mu_C-\mu_H)}{(p_C+p_H)^2} \left( p_H \frac{dp_C}{dt} - p_C \frac{dp_H}{dt} \right)
\right] = 0\,,
\label{eq:volmassdtn2}
\end{align}
where $\mu_C$ and $p_C$ are the molar mass and partial pressure of $C$, respectively, and similarly for $H$.  Recall that quantities with index $v$ relate to the volatile in question, either C or H.  If we set the molar mass of the two species to be identical then the first term in the square brackets drops out, recovering the expected evolution equation for an atmosphere consisting of a single volatile.  However, this first term should not be neglected if the two volatiles have different molar masses.  Furthermore, this term provides a non-linear coupling between all of the volatiles that are present in the system.  Within the time stepper, we solve a system of coupled differential equations to obtain the time updates of the volatile abundances in the melt phase (i.e. $d X_v/dt$).

For the initial condition, it is usually most appropriate to define an initial volatile concentration in the mantle $X_v^0$.  For a mantle that is fully molten at time zero, $M^l=M^m$ and $M^s=0$, which simplifies Eq.~\ref{eq:volmass}.  However, due to coupling between the volatiles through $\bar{\mu}$, as well as potential non-linearities introduced by the modified Henry's law (i.e. when $\beta_v \neq 1$ in Eq.~\ref{eq:partial}), the resulting set of equations are still solved numerically.
\end{appendix}

\end{document}